\documentclass[useAMS,usenatbib]{mn2e}
\usepackage{times}
 \usepackage{txfonts} 
 \usepackage{graphicx} 
 \usepackage{mathabx}
\usepackage{aas_macros}
\usepackage{hyperref}
\usepackage{array}
\usepackage{booktabs}
\usepackage{lscape}
\usepackage{color}
 \usepackage{natbib} 
 
\newcommand{\Ha}{H$\alpha$\,}

\newcommand{\logMs}{$\log{(\rmn{M_{*}/M_{\sun}})}$}

\DeclareRobustCommand{\ion}[2]{%
\relax\ifmmode
\ifx\testbx\f@series
{\mathbf{#1\,\mathsc{#2}}}\else
{\mathrm{#1\,\mathsc{#2}}}\fi
\else\textup{#1\,{\mdseries\textsc{#2}}}%
\fi}

\title[Morpho-kinematics of z$\sim$ 1 galaxies]{Morpho-kinematics of z$\sim$1 galaxies probe the hierarchical scenario }

\author[M. Rodrigues et al.]
  {M.~Rodrigues$^{1}$\thanks{E-mail:myriam.rodrigues@obspm.fr}, 
F.~Hammer$^{1}$, H.~Flores$^{1}$, M.~Puech$^{1}$, E.~Athanassoula$^{2}$\\ 
$^{1}$GEPI , Observatoire de Paris, CNRS, University Paris Diderot, PSL Research University ; 5 Place Jules Janssen,  92195 Meudon, France\\
$^{2}$Laboratoire d'Astrophysique de Marseille (LAM), UMR7326, CNRS/Aix Marseille Universit\'{e}, Technop\^{o}le de Marseille-Etoile,\\
}

\date{Accepted in MNRAS: 23/10/2016}

\begin{document}

\pagerange{\pageref{firstpage}--\pageref{lastpage}} \pubyear{2016}

\maketitle

\label{firstpage}

\begin{abstract}

We have studied a representative sample of intermediate-mass galaxies at z$\sim$1, observed by the kinematic survey KMOS$^{3D}$. We have re-estimated the kinematical parameters from the published kinematic maps and analysed photometric data from HST to measure optical disk inclinations and PAs.  We find that only half of  the z$\sim$ 1 galaxies show kinematic properties consistent with rotating disks, using the same classification scheme than that adopted by the KMOS$^{3D}$ team. Because merger orbital motions can also brought rotation, we have also analysed galaxy morphologies from the available HST imagery. Combining these results to those from kinematics, it leads to a full morpho-kinematic classification. 

To test the robustness of the latter for disentangling isolated disks from mergers, we confronted the results with an analysis of pairs from the open-grism redshift survey 3D-HST. All galaxies found in pairs are affected by either kinematic and/or morphological perturbations. Conversely, all galaxies classified as virialized spirals are found to be isolated. A significant fraction (one fourth) of rotating disks classified from kinematics by the KMOS$^{3D}$ team are found in pairs, which further supports the need for a morpho-kinematic classification.

It results that only one third of z$\sim$1 galaxies are isolated and virialized spirals, while 58\% of them are likely involved in a merger sequence, from first approach to disk rebuilding. The later fraction is in good agreement with the results of semi-empirical $\Lambda$CDM models, supporting a merger-dominated hierarchical scenario as being the main driver of galaxy formation at least during the last 8 billion years.

\end{abstract}


\section{Introduction}
\label{intro}
According to the hierarchical scenario, galaxy interactions play a key
role in the evolution of their morphologies and in how they accrete
mass  \citep[e.g.][]{1972ApJ...178..623T,
  1978MNRAS.183..341W,1988ApJ...331..699B}. The observed properties of
many distant galaxies may seem however to be inconsistent with such
interactions, as galaxies exhibit instead gas-rich, clumpy, and
extended rotating disks not dominated by spheroids
\citep[e.g.][]{2006Natur.442..786G, 2006ApJ...645.1062F,
  2009Natur.457..388G}. \citet{2006MNRAS.368....2D} 
suggested an alternative process to the merger-dominated hierarchical scenario. They proposed that the main mode of galaxy formation was due to mass accretion by cold streams from the IGM, for which simulations showed that it is a dominant mechanism for large mass galaxies at high-z \citep[see, e.g.][]{2011MNRAS.417.2982F}.

This mode of formation is however challenged by recent
cosmological simulations \citep[see, e.g.][]{2011MNRAS.416.2802F, 2012MNRAS.425.2027K, 2011ApJ...742...76G,
  2013MNRAS.434.3142A, Vogelsberger14}, which indicate that realistic disks may also form out of
gas-rich major mergers. More recently, idealized simulations of major
mergers have shown in detail that the corresponding remnants can be
spiral galaxies, having the same properties as local
spirals. Comparisons between simulations and local spirals include
rotation curves, projected stellar density profiles, classical bulge
to total stellar mass ratios and in particular, the detailed
morphology of structures and substructures, such as thin and thick disks, all three types of bulges (classical, boxy/peanut and disky pseudobulges), bars, spirals, rings, etc. \citep{2016arXiv160203189A}. Furthermore, discrepancies in hydrodynamical solvers \citep{2010ARA&A..48..391S} suggest the possibility for a significantly reduced contribution from cold flows \citep[see, e.g.][]{2013MNRAS.429.3353N}.

In light of these new inputs, it may be opportune to re-investigate how observations and their interpretations can allow distinguishing between these scenarios and thus refine the key constraints on galaxy formation. Both galaxy formation modes require galaxies to have been significantly enriched in gas in the past, which is consistent with the increase of galaxy velocity dispersions with redshift \citep[][]{2007A&A...466...83P,2009Natur.457..388G,2014ApJ...790...89K} as well as with the so-called star formation - stellar mass relation \citep[][]{2014MNRAS.443L..49P}. The two galaxy formation modes mainly differ by their predictions on the causal relation between distant galaxies and their descendants in the Local Universe. Despite high velocity dispersions, turbulent gas-rich disks are expected to
preserve most aspects of isolated, virialized disks (shapes and dynamics) over 8 to 10 Gyr. Conversely, should a galaxy have
experienced one major merger during that time, morphologies and kinematics are expected to have been altered for a time equal to the
merger duration, i.e., approximately 2 to 4 Gyr \citep[][]{2009A&A...507.1313H, 2010MNRAS.404..575L, 2012ApJ...753..128P}. 

By measuring to the internal kinematics of galaxies, 3D spectroscopy allow us to probe directly their dynamical state, providing strong constrains on galaxy formation scenarios.  Kinematic surveys span now a wide redshift range, e.g: CALIFA at $z\sim0$ \citep{2013A&A...549A..87H}, IMAGES at $z\sim0.6$ \citep{2006A&A...455..107F,2008A&A...477..789Y}, MASSIV at $z\sim1$ \citep{2010MNRAS.401.2113E}, KMOS$^{3D}$ at $z\sim1-2$ \citep[][hereafter W15]{2015ApJ...799..209W}, SINS and OSIRIS at $z\sim2$ \citep{2006ApJ...645.1062F,2007ApJ...669..929L}, and AMAZE/LSD at z$\sim3$ \citep{2011A&A...528A..88G}. Several of these studies have investigated the evolution of \textit{rotationally-supported} systems and found very high fractions, up to 83-93\% at z$\sim$1.0. However, these yield little observational constraints on galaxy formation because a large fraction of galaxies involved in mergers are also expected to be \textit{rotationally-supported} \citep{2012A&A...542A..54B, 2015ApJ...803...62H}. Galaxies undergoing violent merging processes do not necessarily exhibit highly asymmetrical kinematics in their star-forming gas. Interacting galaxies near their apocenters after first passage could also have kinematic properties apparently similar to those of isolated spirals (see \citealt{2015ApJ...803...62H} and references therein).  Additional indicators such as morphology as traced by stars is required to further identify relaxed spirals from ongoing mergers.\\

In this paper we aim to define a methodology that can robustly identify virialized and isolated disks from galaxies involved in a merging sequence. It assumes that distant galaxies can be classified with the same scheme as used for local galaxies, in terms of dynamical and morphological properties.  An isolated disk is assumed to be virialized, hence easily recognizable as a simple dynamical system with a large number of remarkable geometrical signatures. We have used the $z\sim1$ galaxy sample from KMOS$^{3D}$ (W15, first year sample), which is described in Section 2. Section 3 describes the methodology used to extract the main kinematical and morphological parameters used in classification.  In Section 4, we compute the fraction of rotating disks and non-rotating disks at z$\sim1$ using the classification scheme of W15. We also introduce the morpho-kinematic analyses and provide the fraction of isolated virialized disks. In Section 5, we investigate the concordance between the morpho-kinematic classification and the fraction of galaxies found in interaction based on open grism spectroscopy (3D-HST), allowing us to robustly assess the observed major merger rate. Finally, in Section 6 we discuss the limitations of the kinematic and morpho-kinematic classifications to properly identify isolated disk from ongoing merger. Throughout this work, we adopt $H_0$=70~kms$^{-1}$~Mpc$^{-1}$, $\Omega_M=0.3$ and $\Omega_{\Lambda}= 0.7$. 


\section{Sample selection and methodology}
\label{Framework}

\subsection{Sample and data description}
\label{Sample}
Recent investigations by W15 and \citet{2016MNRAS.457.1888S} have made use of the VLT/KMOS instrument to gather significant samples of spatially resolved kinematics of high-z galaxies in a mass range ($\rmn{M_{*}}\gid$ $10^{10}$ $\rmn{M_{\sun}}$) that is consistent with them being progenitors of Milky Way-mass galaxies. These samples therefore offer the unique opportunity to test both formation scenarios. In this work we make use of the first released sample from the KMOS$^{3D}$ survey, published by W15. The KMOS$^{3D}$ sample was defined in three redshift ranges using a single criterion on the Ks magnitude, Ks$<$ 23. Objects were gathered from the 3D-HST Treasury Survey, which is to date the most complete redshift catalogue in cosmological fields. According to W15, the KMOS$^{3D}$ sample is representative of Milky Way mass galaxies. The KMOS$^{3D}$ sample also has the advantage to have published velocity and dispersion maps, together with archived multi-band imagery from the Cosmic Assembly Near-infrared Deep Extragalatic Legacy Survey (CANDELS). Because CANDELS imagery is not deep enough to probe disk galaxies up to their optical radius beyond z$\sim$1 (see Section \ref{Morphological_Parameters}), we decided to concentrate the analysis on the $z\sim1$ sample from KMOS$^{3D}$. \\
 
\begin{figure} 
\centering
\resizebox{\hsize}{!}{\includegraphics{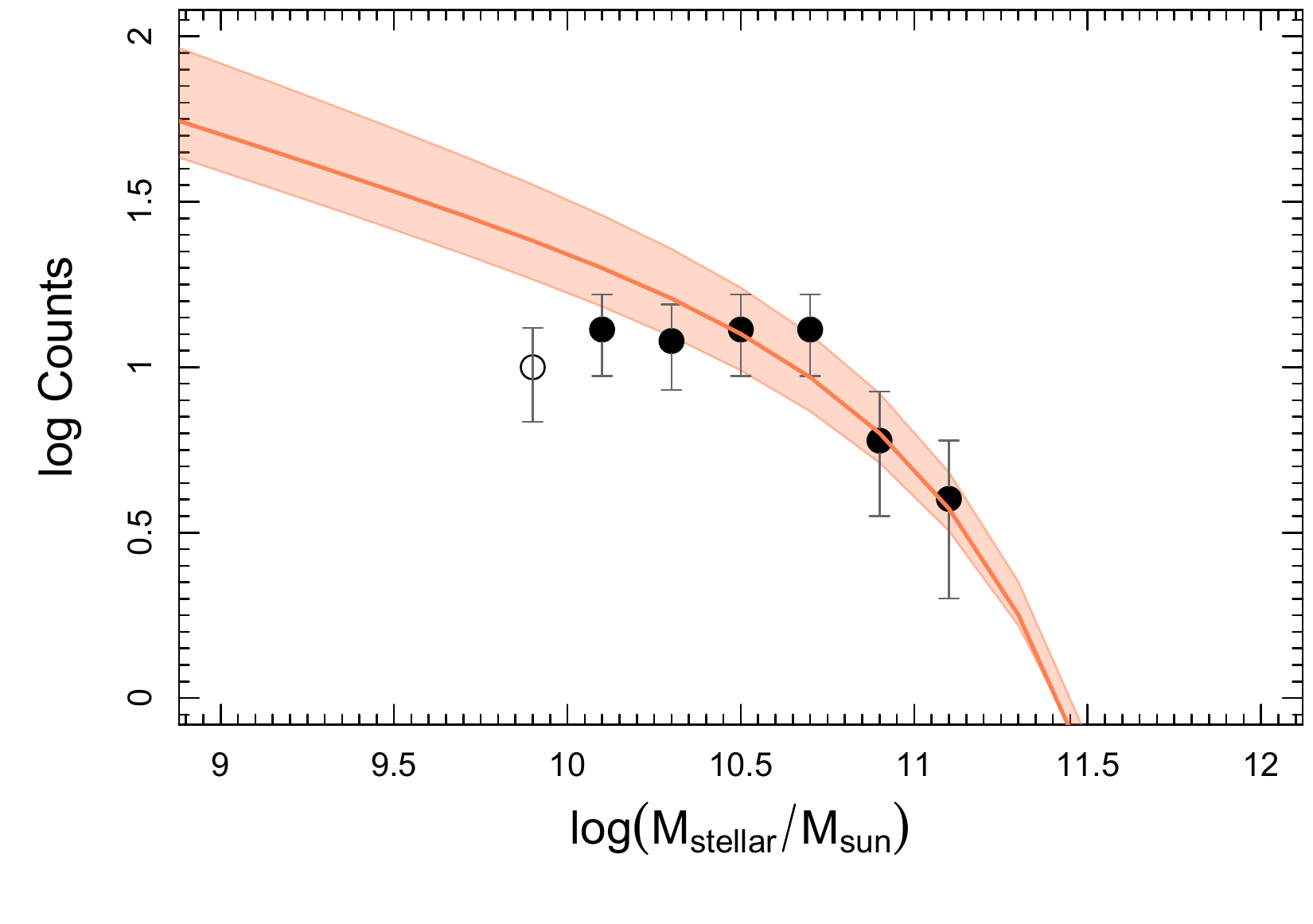}}
\caption{{\small  Number counts in function of stellar mass for the 72 z$\sim$ 1 galaxies from W15, in logarithmic scale. The red line is the stellar mass function at 0.75$<z<$1.25 from \citealt{2014ApJ...783...85T} and the orange area delimits its uncertainties. It evidences that the sample is representative down to \logMs $>$10 (filled black symbols). The open symbol indicates the bin in stellar mass for which the sample is under-representative of the $z\sim1$ population.  }}
\label{Figure_sample}
\end{figure}

This work is based on the $0.7<z<1.1$ galaxies from the KMOS$^{3D}$ first year sample, described in W15. During the first year, 106 z$\sim$1 galaxies were observed at spectral resolution $R=3400$ in the YJ band, with a typical seeing of 0.6$\arcsec$. \Ha emission was detected in 85 galaxies, but only 72 had \Ha emission extending beyond one resolution element. W15 have identified rotationally-supported systems among these 72 galaxies using a series of increasingly stricter criteria. The stellar mass distribution of the 72 z$\sim$1 galaxies from the KMOS$^{3D}$ first year release is presented in Figure \ref{Figure_sample}. The comparison to the stellar mass function \citep{2014ApJ...783...85T} shows that the sample is representative of z$\sim$ 1 galaxies with $\rmn{M_{*}}\gid$ $10^{10}$ $\rmn{M_{\sun}}$. Such a mass range for z$\sim$ 1 galaxies is well consistent for them to be progenitors of Milky Way-mass galaxies ($\rmn{M_{*}}=$ 5.5$\times10^{10}$ $\rmn{M_{\sun}}$, see \citealt{2007ApJ...662..322H}).\\

Annex of W15 provides kinematic maps for a sub-sample of 42 z$\sim1.0$ galaxies, classified as being high S/N disk according to W15 criteria \#1-3 (see a more complete description in section \ref{kine_class}). We restricted our working sample to these 42 high S/N `\textit{disk-like}' galaxies. According to the W15 classification, the remaining 30 objects were split between 14 \textit{disk-like} galaxies with low S/N, while 16 are non-disks. The unavailability of a published object list in W15 required us to identify these 30 objects not by their kinematic maps but relying on their morphology instead. To do this, we compared the colour stamps of the 72 galaxies shown in W15 (Figure 4) with the colour images extracted for all observed objects. 23 galaxies were successfully identified\footnote{We found the same classification compared to W15 for their 16 non-disks galaxies. These objects were straightforward to distinguish from low S/N disks because of their complex kinematics, compact morphologies, and/or obvious merger features near fusion.}, but we failed to secure the identification of the other 7 galaxies. In the course of our investigations, we also identified one galaxy (3D-GS3-23553) that resembles an edge-on spiral with emission on only one side of the disk (see also Figure A1 of W15). Given the associated uncertainty in defining the kinematic parameters, we preferred to discard this object from the W15 sample of high S/N disks (resulting in 41 instead of 42 galaxies), and to consider it as a low S/N disk instead (now 15 instead of 14). The number (16) of non-disk galaxies remained unchanged. \\

We made use of the kinematic maps, velocity and dispersion profiles and intensities as published in the Annex of W15 as well as their corresponding uncertainties. We have assumed that the data reduction and kinematic maps were extracted accurately by W15. The range of values for each map and their associated colour bar scale were recovered as following:
\begin{itemize}
\item for the velocity maps, we have used the minimum and maximum velocities from the velocity profile plots;
\item for the dispersion maps, we have used the minimum and maximal scale from the dispersion profile axis.
\end{itemize}
The values from the profiles plots were measured using a digitizer software ({\sc GraphClick}). All the spaxels in the maps shown by W15 have S/N$>$5. The kinematic parameters were then estimated visually from the maps (see description in section \ref{Kinematical_Parameters}). To test the accuracy of the kinematic maps provided by W15, we re-extracted them for 10 randomly selected galaxies from the W15 high S/N `\textit{disk-like}' (25\% of the sample) and found a good agreement. Only one galaxy had kinematic maps not consistent with that of W15, which would lead to a change on the kinematical analysis (it shows a perturbed rotation instead of being a rotating disk). The difference has a statistical impact smaller than the Poisson uncertainty for such a modest sample, then confirming the quality of the data, of the reduction, and of the kinematical maps of W15. \\

The KMOS$^{3D}$ sample was selected from the open grism 3D-HST survey, which has near-IR imagery in $J_{F125W}$ and $H_{F160W}$ bands from CANDELS \citep{2011ApJS..197...36K} and optical imaging from HST/ACS in $V_{F606W}$ and $I_{F814W}$ bands. This paper makes use of the data compilation\footnote{ 3D-HST is available at \url{ http://3dhst.research.yale.edu/}} from the 3D-HST team \citep{2014ApJS..214...24S}. All ACS images as well as the WFC3 F125W and F140W images have been PSF-convolved to the value of the WFC3/F160W spatial resolution and pixel scales were homogenized to $0.06 \arcsec$ for all the dataset. \\

The stellar mass and UV+IR star formation rate estimates were
extracted from the \citet{2014ApJS..214...24S} and
\citet{2014ApJ...795..104W} catalogues, respectively. The derivation
of both quantities assumes a \citet{2003PASP..115..763C} IMF.

 \section{Classification methodology}
Both the morphological and kinematical classifications were performed independently by three co-authors (FH, HF and MP). The disagreements between the classifiers at the first iteration were discussed individually and resolved. Figure \ref{Figure_high_SN_disk} in Appendix shows for each object, from left to right: : (1) the colour image in
i+J+H-bands (equivalent to local $g$-$r$-$i$ rest-frame colour images), on which are superimposed the optical centre (pink cross), PA (pink line), kinematic centre and PA (respectively cyan cross and cyan dashed line), and the peaks of dispersion (regions delimited by a cyan square); (2) The colour maps produced from the $I_{F814W}$-$J_{F125W}$ images following a S/N weighting scheme that is independent of the colour, which is described in \citet{2004A&A...421..847Z};  (3) the residual map from the 2D decomposition using {\sc galfit}; (4) the decomposed light profile along the major axis in surface brightness units.  

 \subsection{Morphological analysis}
\subsubsection{Morphological parameters}
 \label{Morphological_Parameters}
The morphological PA and b/a ratio were measured from the combination of the two reddest bands ($J_{F125W}$ and $H_{F160W}$) from CANDELS. Their combination samples  $\sim$ 7000\AA~ at rest-frame, which provides a reliable tracer of the distribution of old stars in the optical disk. We used the {\sc iraf} task {\sc ellipse} to fit elliptical isophotes and build radial surface brightness profiles. Neighbouring galaxies around each object were first detected with {\sc sextractor}, and replaced by the mean background level. PA and $b/a$ were measured at the outermost isophote ($1\sigma$ level over the background\footnote{25.7/26./25.9 mag/arcsec$^2$ for respectively COSMOS, GOODS and UDS fields}) and error bars were estimated as the quadratic combination of that provided by the {\sc ellipse} task, and the difference in PA (or in b/a) between the 1$\sigma$ and 2$\sigma$ background levels. The outermost isophote corresponds to 0.54/0.84/0.6$\times$ the optical radius for respectively COSMOS, GOODS and UDS fields, assuming a typical disk with a central magnitude of $\mu_0$=20.7\,mag/arcsec$^2$. For one object (3D-GS3-21045) the PA measurements was biased due to a strong light contamination from several neighboring objects so the PA and ellipticity were measured at a brighter surface brightness for this galaxy.  
The half light radii ($R_{half}$) were measured in the $J_{F125W}$
band images using the curve-of-growth technique. The inclination was
computed from the $b/a$ parameter assuming a thick disk with q=0.25, i.e. $\cos{i}^2=[(b/a)^2 -
  q^2]/[1-q^2]$. \\

 The 2D surface brightness profiles in $H_{F160W}$ were modelled as a combination of a Sersic model (for the bulge) and an exponential disk. We first modelled the radial surface brightness profiles extracted by {\sc ellipse} with a 1D Sersic light profile (bulge) and/or an exponential disk profile (disk), convolved with the image PSF. From this 1D decomposition we retrieved the number of components and a first-guess value of the Sersic index $n$, effective radius $R_e$ and disk scale length $R_h$, total magnitude, PA and $b/a$ ratio of each components. We then modelled the 2D light profile using the {\sc galfit} software \citep{2002AJ....124..266P}, with 1D models as first guesses. All the structural parameters were left free during the fitting process. We used the modelled PSF of WFC3/HST generated by the TinyTim software \citep{2011SPIE.8127E..0JK}. The residuals map from the 2D decomposition and the decomposed light profile along the major axis are shown for each object in Figure \ref{Figure_high_SN_disk}. Morphological parameters - $R_{half}$, $PA_\rmn{opt}$, $b/a$ and B/T from the 2D decomposition - are given in Table \ref{Table_param}.  \\

 \subsubsection{Morphological classification}
\label{Morphological Class}

We classified morphologically the 41 galaxies into Spiral (Sp) and Peculiar (Pec), using a set of physical parameters derived from optical rest-frame imagery (see above): $R_{half}$ to identify compact galaxies; $B/T$ to separate between bulge-dominated and disk-dominated galaxies; colour maps to identify star-forming and/or dusty regions, and detect colour asymmetries; residual maps from 2D surface brightness decomposition to discriminate between spiral galaxies having symmetrical residuals (spiral arms, rings or bars) from disrupted galaxies having asymmetric residuals such as tidal tails, see Figure~\ref{Figure_high_SN_disk}. 

The morphological classification was performed by following the methodology of \citet{2010A&A...509A..78D}, formalised by a decision-tree. Peculiar galaxies include galaxies that are either compact or blue-cored, or with asymmetric behaviours, which can be identified from their colour and residual maps (see Figure \ref{Figure_high_SN_disk}). Conversely, spiral galaxies are not blue-cored and present symmetric features (arms or rings, bars) in the residuals maps. The morphological classification of each galaxy is given in Table \ref{Table_param}. The level of agreement between classifiers after the first iteration is given by a confidence flag:  0 = all classifiers agree; 1 = when classifiers disagree. The initial disagreement between classifiers was $\sim 12\%$.
\subsubsection{On the measurement of the viewing angles: bias introduced by single Sersic fitting}
\label{PA_measures}
\begin{figure}
\centering
 \resizebox{\hsize}{!}{\includegraphics{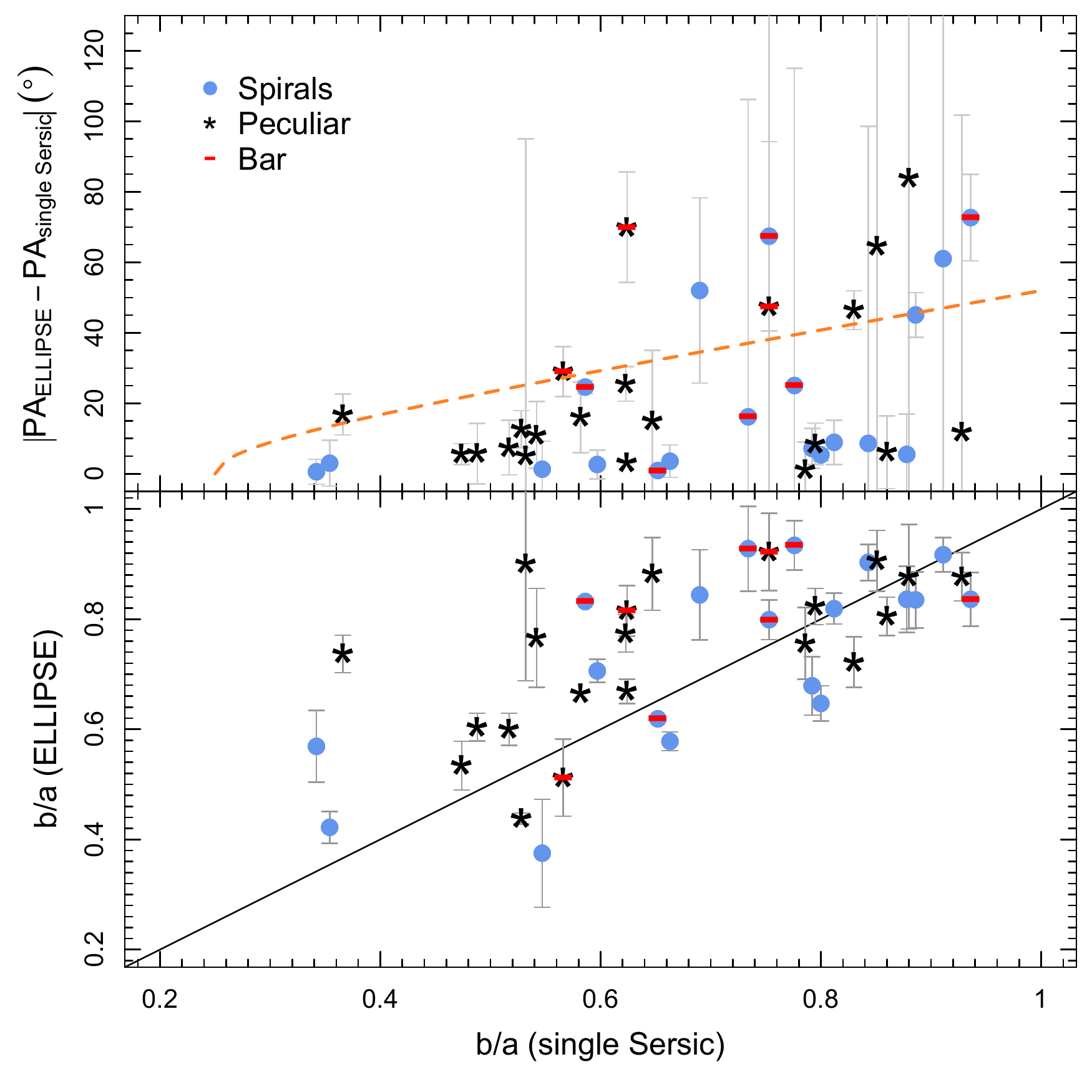}}
\caption{{\small Comparison between morphological parameters estimated by a single Sersic fitting \citep[from the CANDELS catalogue, see][]{2012ApJS..203...24V} and those measured at the outer isophotes by {\sc ellipse} (this work).  \textbf{Upper panel}: Differences between $\rmn{PA_{single\,Sersic}}$ and $\rmn{PA_{ELLIPSE}}$  as a function of the b/a ratio. The morphological classification results are symbolised as follows (see also Section~\ref{Morphological Class}): blue circles (spirals), black stars (peculiars), while a superposed red bar indicates the presence of a strong bar.  The orange dashed line represents the effect of projection of a misalignment of 60$^\degree$ between a bar and a disk. \textbf{Bottom panel}: It shows a comparison between $b/a$ estimated from a single Sersic model and  measured using {\sc ellipse} (this work). }}
\label{Compare_Candels_morpho}
\end{figure}

\begin{figure}
\centering
  \resizebox{\hsize}{!}{\includegraphics{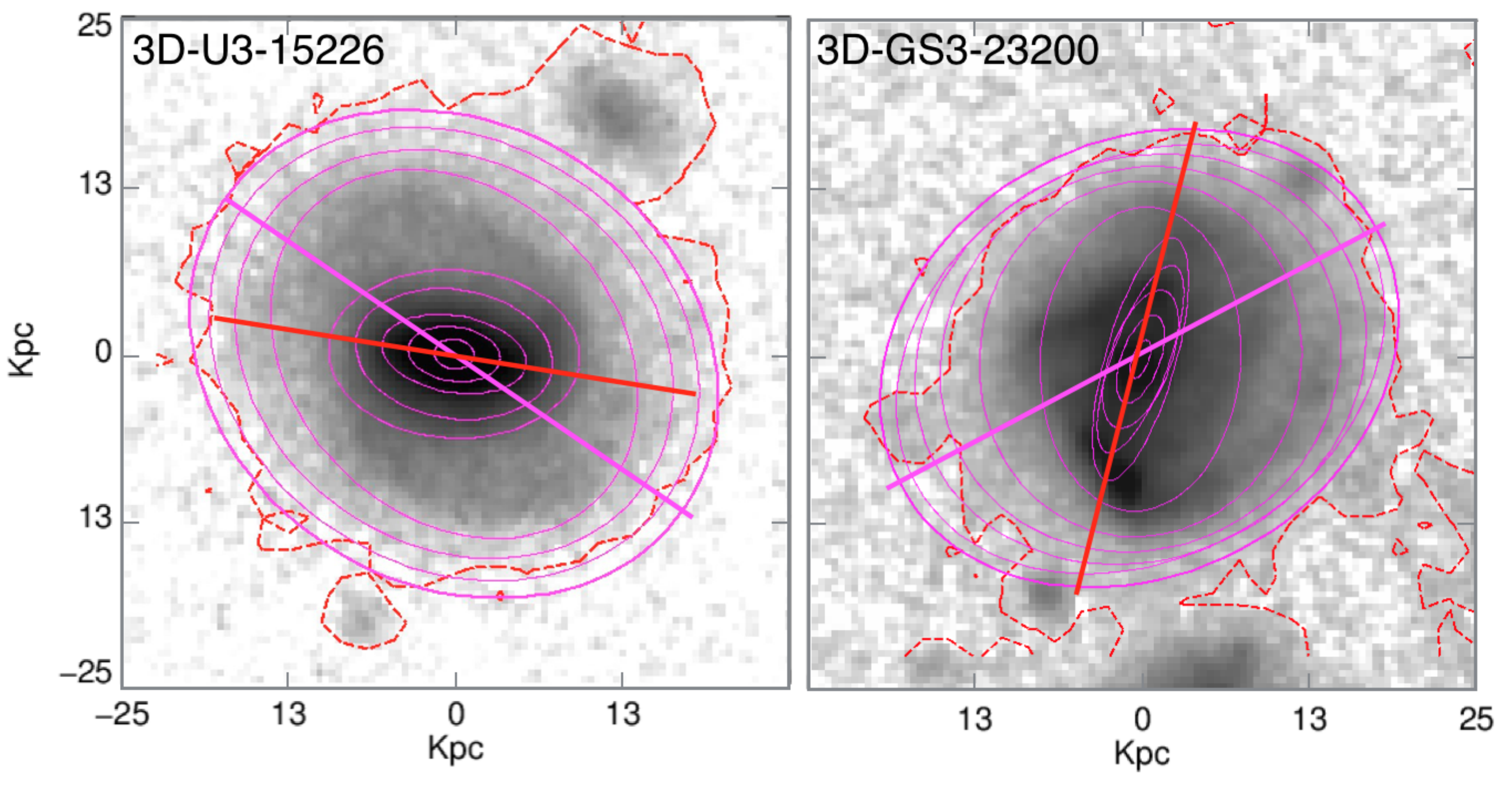}}
\caption{{\small Two examples of barred galaxies for which the morphological PA estimated from a single Sersic component (red line) leading to a PA along the prominent bar that misses the actual galaxy PA.  }}
\label{PA_difference_bars}
\end{figure}

The $PA_\rmn{opt}$ and $b/a$ used by W15 were extracted from \citet{2012ApJS..203...24V} who used {\sc galfit} to model light profiles using a single Sersic component. In Figure~\ref{Compare_Candels_morpho}, we compared the values of PA and b/a estimated by the single Sersic modelling with those measured in the outer isophotes using {\sc ellipse} (this work). The upper panels show the difference $|\rmn{PA_{ELLIPSE}}$ - $\rmn{PA_{single\,Sersic}}|$ as a function of $b/a$ from a single Sersic fit. The mean difference over the 41 galaxies is 27$^\degree$. We have investigated if the discrepancy could be related to specific morphological features. Barred galaxies and galaxies with peculiar morphologies systematically have higher PA and b/a discrepancies: barred, peculiar and non barred spiral galaxies show an average $|\rmn{PA_{ELLIPSE}}$ - $\rmn{PA_{single\,Sersic}}|$ of 25, 36 and 18 $^\degree$, respectively. Figure~\ref{PA_difference_bars} shows two examples of barred galaxies with discrepant PA measurements. The PAs estimated by a single Sersic (red axis) follow the bar axis and not the PA of the disk (pink axis). The reason is that a light profile decomposition using such a simplified model is systematically weighted toward brighter regions, e.g. bar or peculiar features. This is corroborated by the bottom panel of Figure~\ref{Compare_Candels_morpho}: in the presence of a bar (red symbols), a single Sersic fits is weighted toward the small b/a value for a bar component. The lack of b/a $<$ 0.6 objects having large $|\rmn{PA_{ELLIPSE}}$ - $\rmn{PA_{single\,Sersic}}|$ values may result from projection effects (see also a similar explanation in section \ref{Fraction_Rot}) since a variation of PA as a function of radius is much more difficult to detect in inclined objects. This is illustrated by the dashed orange line in Figure \ref{Compare_Candels_morpho}, which shows the observed Delta PA as a function of b/a for an intrinsic misalignment of 60$^\degree$ between a bar and a disk. At high inclination, the light profile is mainly dominated by the disk, and both single Sersic and {\sc Ellipse} methods will give consistent measurements. On the contrary, at lower inclination, the variation of PA with radius can be easily detected in the light profile. In this case, single Sersic estimates will be biased toward the PA of the internal isophotes.

In the next section, morphological and kinematic PAs are compared to identify non-virialized objects, which are expected to show strong mismatches between ionised gas and stellar distributions. Conversely, in a virialized disk the stellar component should be regular up to the edge of the disk, and so the ionised gas. While the kinematic PA is not affected by the presence of a bar or bright central features, the morphological PA measured in the inner isophotes is. As such, the PA measured in the outer isophotes provides a more robust measure of the stellar disk PA. Given the peculiar features dominating the morphologies of distant galaxies (e.g. bars, clumps, tidal tails), we highlight that the choice of the methodology used to measure $\rmn{PA_{opt}}$ is a crucial point for establishing a proper kinematic classification.

 \subsection{Kinematical classification}
\label{kine_class}

 \subsubsection{Kinematical Parameters }
 \label{Kinematical_Parameters}
For each galaxy we measured the kinematic PA ($PA_{kin}$), kinematic centre
($c_\rmn{kin}$), and position of the sigma peak directly from the
velocity-fields and sigma maps published in W15 (\Ha line detected with SNR$>$5, see Figure~\ref{Figure_high_SN_disk} in Appendix A). 

The kinematic position angle was set to the axis defined by the position of \textit{max} and \textit{min} velocities. The uncertainty on $PA_{kin}$ is evaluated to be 10$^\degree$. The kinematic centre ($c_\rmn{kin}$) was measured at the average position between the \textit{max} and \textit{min} velocities observed along the $PA_{kin}$. We considered that a sigma peak is detected when an increase in dispersion is observed over several contiguous pixels, i.e., approximately the same number of pixels covered by the PSF. An offset sigma peak is detected when it shows a significant increase of the dispersion\footnote{The instrumental resolution of KMOS$^{3D}$ observations is $\sigma_{instr}=27-46\,km/s$. To investigate how far below the spectral resolution limit dispersion can be measured, W15 used a suite of model emission spectra (in their section 3.2). They found that dispersions between $\sigma_{instr}-0.3\sigma_{instr}$ can be recovered with a 30-60\% uncertainty.} ($>$10\,km/s) when compared to the value at the centre of rotation. \\

\begin{figure} 
\centering
 \resizebox{\hsize}{!}{\includegraphics{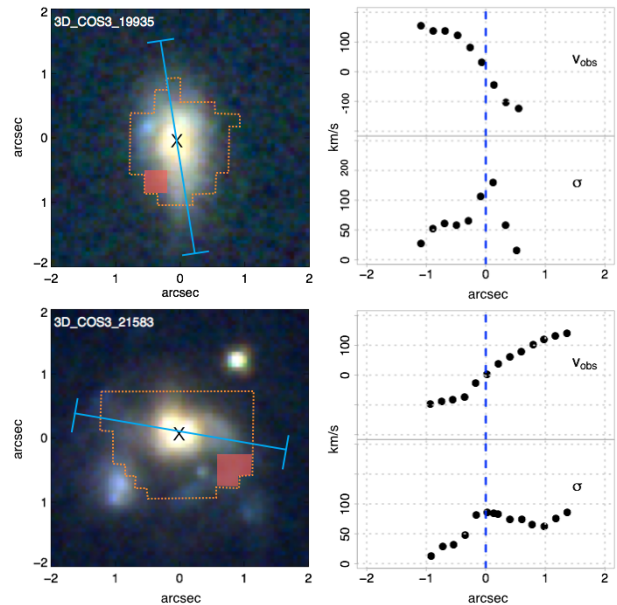}}
\caption{{\small  Measurement of the $\sigma$-peak position in objects 3D-COS3-19935 (upper panels) and 3D-COS3-21583 (lower panels). The left panels show the i- J- H- colour image. The orange dashed line delimits the detected H$\alpha$ emission. The red area corresponds to the position of the sigma peak. The blue line indicates the kinematic axis from which the velocity and velocity dispersion profiles (right panels) have been extracted. The velocity and dispersion profiles were extracted from the right panels of Figure A1 in W15. The dashed blue line corresponds to the centre of rotation indicated in W15 (the black vertical dot-dashed line in the W15 figures). For clarity, we did not over-plot the other lines provided by W15. }}
\label{Figure_sigma_peak}
\end{figure}

We defined $r_{kin}$ as the distance between the kinematic centre and the peak of velocity dispersion. It was measured as being the distance between the barycentre of sigma values in the sigma peak and the kinematic centre $c_\rmn{kin}$, following the method established by \citet{2006A&A...455..107F}. This methodology takes advantage of the full information provided by 3D observations searching for all the peaks of velocity dispersion within the IFU. In contrast, W15 used only the velocity and sigma profiles along the major axis (slit-like observations), which can lead to drastically different results. This is illustrated in Figure \ref{Figure_sigma_peak} for 3D-GS3-21583 and 3D-COS3-19935. The left panel shows the extension of the H$\alpha$ emission detected by the IFU (orange doted line) over a colour image. The peak of dispersion is indicated by a red square. The right panels plot the velocity and sigma profile along the kinematic axis (blue axis in the left panel).  While the methodology of \citet{2006A&A...455..107F} detects a significant offset of the sigma peak from the rotation centre (black cross), there is no evidence for an offset from the sigma profile. This last method is only sensitive to local peaks along the kinematic axis.

The distance between $c_\rmn{kin}$ and the optical centre was measured by comparing the continuum images to the $H\alpha$-maps and to the velocity fields. Because continuum images were not available in W15, we reduced the raw KMOS data available in the ESO archive using the Software Package for Astronomical Reduction with KMOS \citep[SPARK][]{2013A&A...558A..56D} and the common SPARK recipes, to extract continuum and \Ha images. Continuum images were obtained by collapsing the cube along the spectral dimension. Strong sky lines residuals were previously masked using a $\sigma$-clipping algorithm. \Ha-images were extracted by summing the pixels in a narrow spectral window delineating the line. We then matched directly the \Ha-images to the \Ha-maps of W15, and measured the offset between the optical centre in the continuum images to the kinematic centre from the $H\alpha$-map.
 
\subsubsection{Criteria to define a rotating disk}
\label{Criteria_kine}

We classified kinematically the sub-sample of 41 high S/N 'disk-like' galaxies, using the same classification scheme of W15, i.e., a galaxy is a rotating disk if it verifies all the following five criteria. The first two criteria select galaxies which are rotationally-supported because:
\begin{description}
\item[(1)]  The velocity map has a single velocity gradient;
\item[(2)]  $V_\rmn{rot}/\sigma _0 >$1, where $\sigma_0$ is the intrinsic disk dispersion, i.e., after deconvolution of projection effects in the dispersion map.
\end{description}

Due to the coarse resolution, distant rotating disks are easily recognizable by the presence of a dispersion peak coincident with the centre of rotation. Indeed, at low resolution, the gradient of velocity at the centre of rotation is unresolved and is detected as a peak in the dispersion map. It leads to the $\sigma$-centering criterion \citep{2006A&A...455..107F}, which defines whether the rotation is perturbed or not:
\begin{description}
\item[(3)] There is a sigma-peak coinciding with the centre of rotation within $r_{kin}\sim$1.6 pixels ($\sim0.25\arcsec$). This criterion is not taken into account for: (i) galaxies that are almost face-on, for which the peak of dispersion is less strong due to the flatter velocity gradient; (ii) galaxies with weak star formation at the centre for which \Ha emission may not be bright enough to generate a sigma peak.
\end{description}

These three kinematic criteria are not sufficient to properly disentangle rotating disks from interacting/merging galaxies. Adding information from high spatial resolution imagery is needed to identify strong misalignments between gas and old stars typically observed during major fusion episodes or outflows \citep{2015A&A...582A..21B}. To do this, the distribution of old stars should be constrained from high spatial resolution imagery at rest-frame wavelengths above 4\,000\AA. The requirement of an equilibrium between the gas and old stars in rotating disks is translated into the fourth and fifth criteria of W15:
\begin{description}
\item[(4)] There is no mismatch between the kinematic and morphological PA, i.e., $\Delta PA=PA_\rmn{kin} - PA_\rmn{morph} < 30^\degree$. In the local universe, non-interactive galaxies have their kinematic and morphological PA aligned within less than 20$^\degree$ \citep{2008MNRAS.390..466E, 2014A&A...568A..70B,2015A&A...582A..21B}. The threshold of 30$^\degree$, assumed in this work, takes into account the large uncertainties on the estimation of both PAs at high redshift. This criterion is not taken into account for nearly face-on galaxies (in practice those with b/a $>$ 0.85). 
\item[(5)] The rotation centre does coincide with the centre of stellar mass by $\Delta c=dist(c_\rmn{kin} - c_\rmn{opt}) < 0.4\arcsec$ ($<$ 2 pixels).
\end{description}

The resulting kinematic class of each galaxy is given in Table \ref{Table_param}, together with the level of agreement between classifiers after first iteration. The initial disagreement between classifiers was $< 10\%$. To facilitate the comparison with previous works, Table~\ref{Table_param} also gives the kinematic class of \citet{2008A&A...477..789Y} for each galaxy: rotating disks (RD) are galaxies satisfying all the criteria above; perturbed rotators (PR), which are rotationally supported systems but not yet virialized, satisfy all above criteria but criterion (3) ; and galaxies with complex kinematics (CK) are disrupted objects, which are rejected either by criteria (1), (4) or (5). \\

\subsection{Fraction of rotating disks and Comparison with W15 kinematic classification}
\label{Fraction_Rot}

From the 41 galaxies classified as \textit{disk-like} by W15 (criteria 1-3), 2 fail criterion (1). 3D-COS3-10248 and GS3-21045 show a velocity field with multiple velocity gradients (see Figure A1 of W15 for comparison). They both show what could resemble an extremely strong warp in their velocity fields, which indeed provides two discrepant kinematic axes. An examination of the mapping extended to low S/N spaxels (see Figure 5 of W15) evidences complex velocity fields with more than 2 extrema. It could be argued that the criterion (1) may depend on a certain subjectivity. However 3D-COS3-10248 and GS3-21045 are furthermore rejected as rotating disks on the basis of pure methodological and reproducible criteria (3 and 4, respectively, see Table~\ref{Table_param}). 

For the 20\% of galaxies that failed criteria (1--3), the major discrepancy with the W15 results comes from criteria (3) for which 6 out of the 41 galaxies were rejected. This is due to the different methodologies used to measure the $\sigma$-peak offset discussed in Section \ref{Kinematical_Parameters}: while our methodology detects peaks over the whole dispersion map, the W15 analysis was restricted to the sigma-peaks lying along the kinematic axis. 

\begin{figure*}
\centering
 \resizebox{\hsize}{!}{\includegraphics{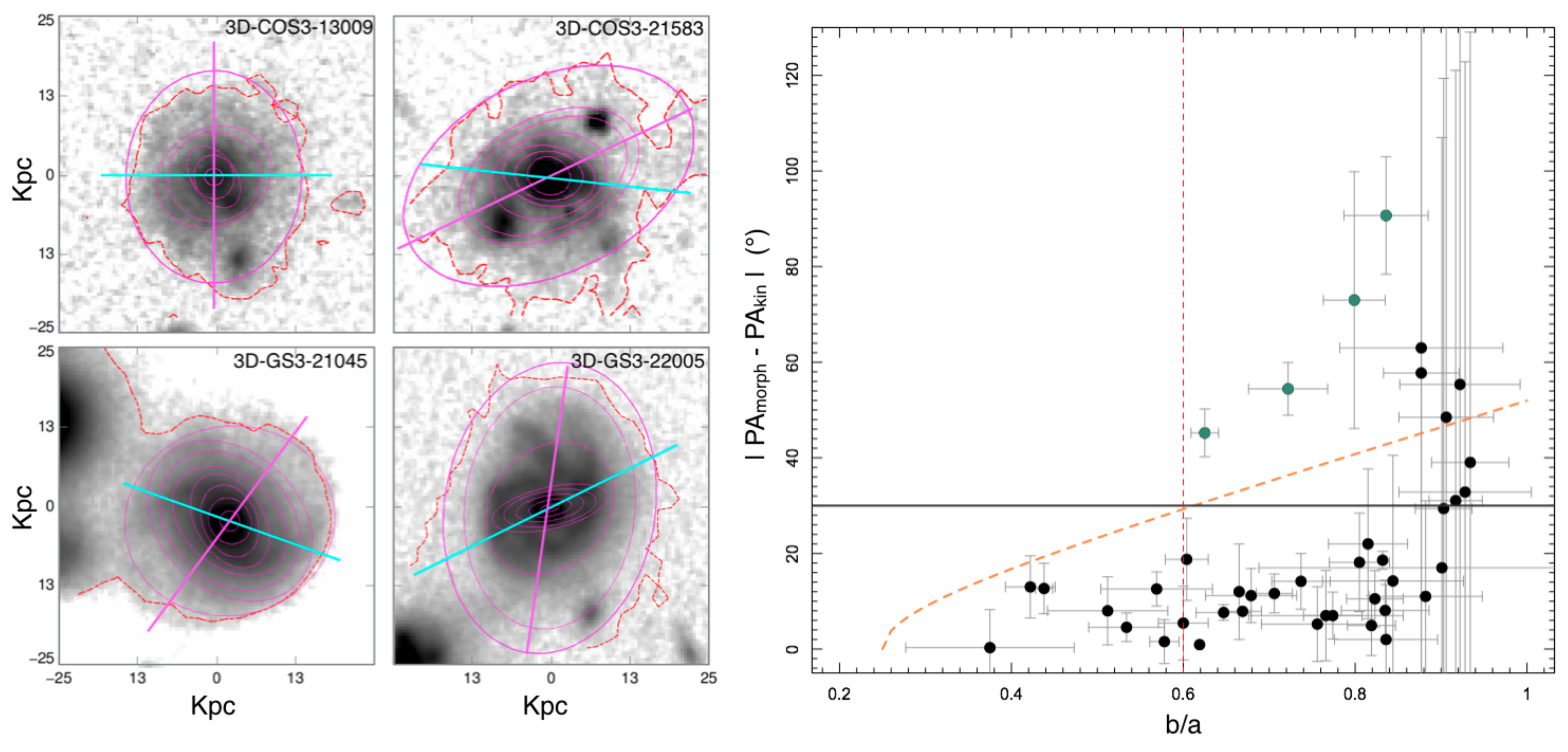}}
\caption{{\small \textbf{Left panels}: J+H images of 3D-COS3-13009, 3D-COS21583, 3D-GS3-21045 and 3D-GS3-22005, that show the strongest misalignments between their morphology and kinematics. The dashed red lines give the outer isophotes,  at 1-$\sigma$ over the background); The pink ellipses are the elliptical isophotes fitted by {\sc ellipse}; The pink and cyan solid straight lines represent the morphological PA measured at the outer isophote and the kinematic PA, respectively. \textbf{Right panel}: Difference between the morphological and kinematic PAs ($\Delta PA$). The black solid line indicates the limit of $\Delta PA>30^\degree$ over which a galaxy is classified as not-rotating. The green symbols are the 4 galaxies with $\Delta PA>30^\degree$ shown in the left panels. The orange dashed line represents the effect of projection of a morpho-kinematic misalignment of 60$^\degree$ in the stellar disk plane (see text). The red vertical dashed line shows the additional constraint on galaxy inclination, $b/a < 0.6$, taken by W15 for criterion (4) on $\Delta PA$.}}
\label{PA_mismatch}
\end{figure*}

Four galaxies exhibit a significant morpho-kinematic PA misalignment  $\Delta \rmn{PA}>30^\degree$, as illustrated in Figure~\ref{PA_mismatch} (left). Only one galaxy was rejected by this sole criterion, while the other three were already rejected because of criteria (1-3). The fact that W15 found no galaxy in their high-S/N 'disk sample' affected by such a misalignment (see their Table 1) is due to their additional requirement to consider misalignments only for galaxies with b/a$<$0.6. In nearly face-on disks, morphological PAs are unconstrained, justifying the need for a cut on b/a. In this work, we used the uncertainties on $\Delta \rmn{PA}$ to verify if criterion (4) is passed, which is equivalent to a cut at b/a$\sim$0.85 (see the points with large error bars on the rightmost side of Figure \ref{PA_mismatch}). For comparison if we had used a cut at b/a$<$0.6 (inclination $>55^\degree$) as chosen by W15, it would have restricted the analysis to only 7 out of 41 galaxies, as illustrated in Figure \ref{PA_mismatch} (right, galaxies on the left of the vertical red dashed line). Moreover, observations at low b/a are subjected to biases that hide the impact of possible interactions. For example, during first passage to fusion, the angular momentum transfer may also lead to radial distortions, i.e., along the disk plane. However, distortions along the disk plane can be almost impossible to detect due to projection effects: if seen with an inclination $i$, the observed $\Delta \rmn{PA_{obs}}$ of an intrinsic difference of $\Delta \rmn{PA_0}$ along the disk plane is given by $\Delta \rmn{PA_{obs}}  = \Delta \rmn{PA_0}  \, \cos{(i)} \, \sin{(\Delta \rmn{PA_0})}$. The orange dashed line in Figure \ref{PA_mismatch} (right) shows how a strong morpho-kinematic misalignment of $\Delta \rmn{PA_0}= 60^\degree$ within the disk plane is seen at a given b/a: for b/a$<$0.6, $\Delta \rmn{PA_{obs}}$ systematically falls under 30$^\degree$. This is a further reason why the W15 study was unable to find significant misalignments. Finally, 4 galaxies were rejected because of criterion (5), in quite good agreement with the W15 classification.

Among the 41 objects classified as \textit{disk-like} galaxies by W15, we found that 80\% successfully passed the same criteria (1--3), and 68\% when assuming criterion (1--5) compared to 89\% found by W15. Table~\ref{Table_kine}, rows 6 to 11, summarises the results of our kinematic classification. 

\setcounter{table}{1}
 \begin{table}
\centering
\caption{ Kinematic classification of the `\textit{disk-like}' galaxies from W15, using the following set of cumulative criteria: (1) Single velocity gradient; (2) $V_{\rmn{obs}}/\sigma_0 >  1$; (3) $\Delta r_{kin}<1.6$~pixels; (4) $\Delta PA<30^\degree$ (for W15 assuming only b/a$<$ 0.6 galaxies); (5) $\Delta c<$2~pixels. N is the total number of  objects in each sample. The top rows provide the classification of W15 (see their table 1) for their 56 `\textit{disk-like}' galaxies. The bottom rows give the classification made in this study for a sub-sample of 41/56 `\textit{disk-like}' galaxies for which the kinematic maps are available from W15. For a criterion (n), the rows 'Number' indicate the number of galaxies that verify criterion (1) to criterion $(n)$; the rows 'Rejected' give the number of galaxies that have been rejected from being rotating disk because of the specific criterion $(n)$;  the row 'Not compliant' gives the number of galaxies that do not verify a given criterion $(n)$. Galaxies could failed several criteria but are only rejected as being disk by a single criterion. }
\begin{tabular}{llrrrrrr}
\toprule
   &Criterion & N& (1) & (2) & (3) & (4) & (5) \\ 
   \midrule
    \multicolumn{4}{l}{\textbf{W15}: `\textit{disk-like}' galaxies} &   &    &    \\ 
&Number  &56&&&  & 56  & 50 \\ 
&Rejected && \multicolumn{2}{r}{}&  & 0   & 6   \\ 
& Fraction (\%)&100\%& \multicolumn{2}{r}{ } &   &100\%   & 89\% \\ 
 \midrule
  \multicolumn{8}{l}{\textbf{This work}: Sub-sample of 41 high-SN `\textit{disk-like}' galaxies from W15} \\ 
 & Number&41&39 &39&33&32 & 28  \\ 
& Rejected&    &2  &0  &6  &1    &4   \\ 
 & Not compliant &&2&0&7&4&5  \\ 
& Fraction (\%)&100\%&95\%&95\% &80\%  &78\%   & 68\% \\ 
\bottomrule
\end{tabular}
\label{Table_kine}
\end{table}

\section{Disk and merger fractions at z$\sim$1}

\subsection{Fraction of rotating disks from kinematic classification}
\label{f_disks}

Despite applying the same kinematic criteria as W15, we have found noticeably different results: only 68\%$\pm$16\% amongst the 41 \textit{'disk-like}' galaxies of W15 are found to be rotating disks (61\%$\pm$16\% if restricting to galaxies above the mass-limit $\rmn{M_{*}}\gid$ $10^{10}$ $\rmn{M_{\sun}}$). We emphasize that, conversely to the W15 suggestion, criteria (4) and (5) are not optional but mandatory conditions to define a relaxed disk. For example, \citet{2014A&A...568A..70B,2015A&A...582A..21B} found that 10\% (52\%) of local spirals (mergers) show $\Delta PA>$22$^\degree$.  These misalignments are indeed found in galaxies with morphological signatures of interaction, mainly between the first and second passages when the angular momentum exchange is expected to be large. It is also the case for the LMC disk that shows $\Delta$PA $\sim$ 45$\degr$ (see, e.g., \citealt{vanderMarel14}), which could be attributed to the recent 200~Myr old collision with the SMC (see \citealt{2015ApJ...813..110H}). In the same way, criterion (5) has to be complied by any virialized, rotationally supported system. The 2 spaxels (0.4$\arcsec$) offset between the old stars and dynamical centres translates into a projected distance $\sim$3.2~kpc, leading to an even larger physical separation when corrected for inclination. This would be equivalent to a virialized Milky Way mass galaxy rotating around an axis located at half the distance between the Sun and the bulge. \\

The 68\% of rotating disks found in the sample of 41 high S/N \textit{'disk-like}' galaxies (see Table~\ref{Table_kine}) finally translates into 53\% of rotating disks when accounting for the full sample of z$\sim$1 galaxies. Table~\ref{Table_fractions} gives the fraction of galaxies of the $z\sim1$ sample verifying disks criteria by W15 (row 1) and by this work (row 2). We have assumed that the properties of the 41 galaxies are representative of the 56 \textit{disk-like} galaxies, and have added into the statistics the 16 \textit{'non-disk'} identified by W15. These results indicate a significantly smaller disk fraction (53\%) than that reported by W15 (78\% and 70\% for criteria 1-3 or 1-5, respectively). The description of individual targets can be found in Annex \ref{Description_target}.

 \begin{table}
\centering
\caption{ Kinematic and (in last column) morpho-kinematic classifications of $z\sim$1 galaxies. The first 5 columns are the kinematic criteria described in section ~\ref{Criteria_kine}. The last column corresponds to adding a criterion based on morphology, leading to the full morpho-kinematic classification (see Section~\ref{Morpho-kine}). $^\dagger$ criteria (1) and (2) are merged in Table 1 of W15.}
\begin{tabular}{lrrrrrrr}
\toprule
  Criteria & 1 & 2 & 3 & 4 & 5&Morpho-kinematics \\ 
   \midrule
 \textbf{W15}  &\multicolumn{2}{c}{93\%$^\dagger$}&78\%  & 78\%   & 70\%  & \\ 
\textbf{This work} &74\%&74\% &63\%  &61\%   & 53\%  &25\% \\ 
\bottomrule
\end{tabular}
\label{Table_fractions}
\end{table}

\subsection{Disk/merger ratio from a morpho-kinematic classification}
\label{Morpho-kine}

Kinematic studies of distant galaxies may be limited when compared to morphological studies based on the HST imagery. For some galaxies, kinematics may indicate virialised systems, while the imaging may still reveal stellar bridges and tails in the outskirts because those have larger characteristic times to reach an equilibrium than bright central regions. Morphology includes also the color information, and strong perturbations of the inner parts could be sometime undetected due to the poor spatial resolution of the kinematic maps. This has prompted us to identify those galaxies having either compact, blue-cored or asymmetrical morphologies (all wrapped into a single category of peculiar galaxies), and to adopt the \textit{morpho-kinematic} method proposed by \citet{2009A&A...507.1313H}. The morphological classification (Section \ref{Morphological Class}) was combined to the kinematic classification (Section \ref{Morphological Class}), leading to 3 morpho-kinematic\footnote{This work uses a slightly different terminology than in \citet{2009A&A...507.1313H} by replacing the term "relaxed" instead of "virialized". The later is preferred because it is more appropriate to describe an isolated disk. While a galaxy is virialized after few disk rotations, relaxation takes more than a Hubble time to achieve. } categories:
\begin{itemize}
\item Virialized, rotating spiral (RS) are galaxies having kinematics consistent with those of rotating disks (RD), and showing the morphology of
a spiral galaxy;
\item Semi-virialized (SV) galaxies possess either a rotating disk kinematics
 and a peculiar morphology or kinematics discrepant
from rotation but a spiral morphology;
\item Non-virialized (NV) systems are galaxies with kinematics discrepant
from a rotational velocity field (CK and PR) and whose morphology is also peculiar.
\end{itemize}

Figure~\ref{Figure_high_SN_disk} and Table 1 give the morpho-kinematic classification for the 41 high S/N disks of W15, which include 13 virialized disks, 15 semi-virialized galaxies with peculiar morphologies, 6 semi-virialized galaxies with peculiar kinematics, and 7 non-virialized systems. As a consequence, we found that only 25\% of $z\sim1$ galaxies can be securely identified as isolated virialized spirals, the rest being at different stages of a merger sequence, in sharp contrast with W15. Restricting the sample to objects in the completeness interval ($\rmn{M_{*}}\gid$ $10^{10}$ $\rmn{M_{\sun}}$), the fraction of isolated spirals at $z\sim$1 is about 31\%.

\section{Testing the robustness of the morpho-kinematic classification }

\subsection{Pair fractions from the 3D-HST survey: companions and major mergers}
\label{pairs}
 
\begin{figure} 
\centering
 \resizebox{\hsize}{!}{\includegraphics{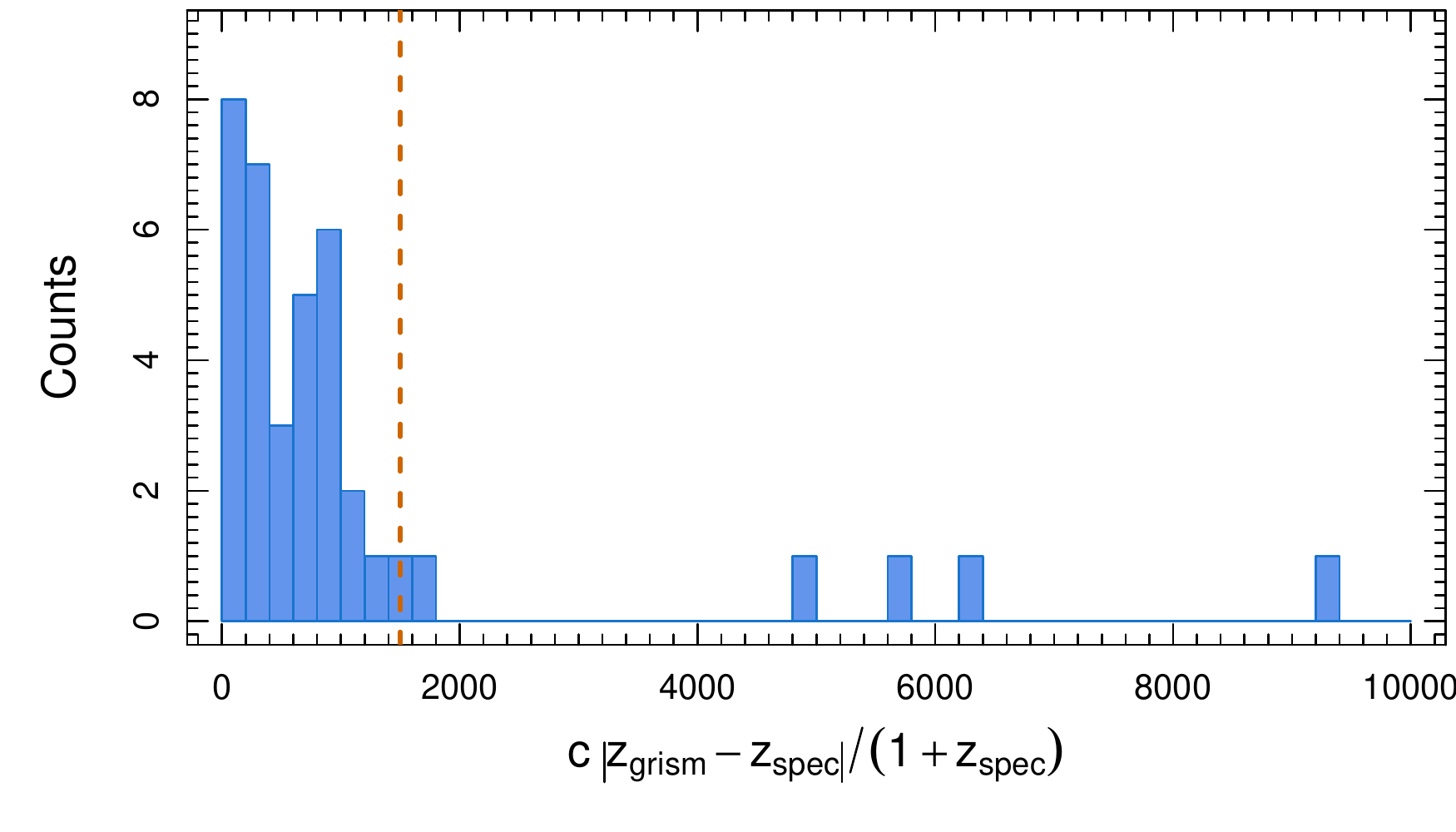}}
\caption{{\small  Comparison between $z_{\rmn{spec}}$ and $z_{\rmn{grism}}$ for  39 galaxies of the high S/N disk sample of W15. The Figure shows the corresponding velocity difference with a mean and $\sigma$ being 1200 and 1975~km\,s$^{-1}$, respectively. The vertical, dashed line at $\Delta V$= 1500~km\,s$^{-1}$ includes 78\% of the sample.  Two of the 41 high S/N galaxies do not have $z_{\rmn{grism}}$ measurements. Notice also that one object has an extremely catastrophic $z_{\rmn{grism}}$ measurement ($\Delta z$ = 0.313).}}
\label{Delta_V_grism}
\end{figure}

To test the robustness of the morpho-kinematic classification in identifying (un)perturbed systems, we investigated the presence of galaxies in pairs. 
We made use of the 3D-HST redshift catalogue by \citet{2015arXiv151002106M}, which is to date the most complete redshift catalogue in the CANDELS field. This catalogue compiles spectroscopic redshifts ($z_{\rmn{spec}}$) from previous spectroscopic surveys, with redshifts obtained by the open grism observations ($z_{\rmn{grism}}$). Most of the available redshifts are based on grism observations, and have much higher uncertainties than spectroscopic redshifts. As shown in Figure~\ref{Delta_V_grism}, about 78\% of our sample has $|z_{spec}$ - $z_{grism}|$/(1+$z_{\rmn{spec}}$) within 1500~km\,s$^{-1}$. This uncertainty is consistent with that estimated by \citet{2012ApJS..200...13B} for a z$\sim$1.3 galaxy with (spatial) FWHM = 0$\arcsec$.5 observed at the grism spectral resolution. In a few cases, grism spectroscopy could lead to catastrophic redshift estimations, with discrepancies well above $\Delta V$= 1500~km\,s$^{-1}$.

\begin{figure*}
\centering
 \resizebox{\hsize}{!}{\includegraphics{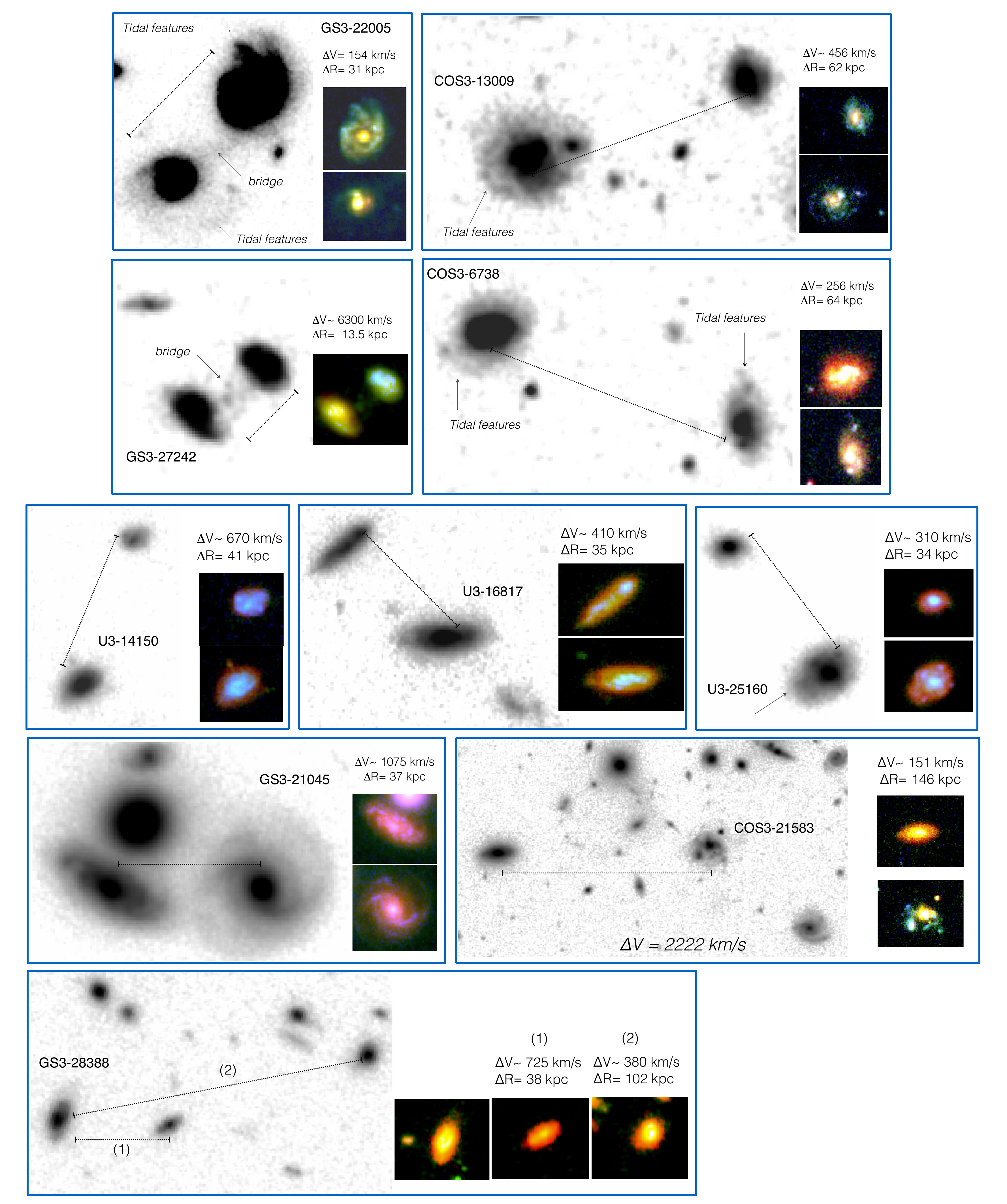}}
\caption{{\small  Galaxy pairs found in the W15 sample of high S/N \textit{disk} galaxies, within $r_{proj}<$150~kpc and $\Delta V< 1500$~km$s^{-1}$, when calculated from $\Delta z_{\rmn{grism}}$ (otherwise $\Delta V< 500$~km$s^{-1}$). In the different panels, $\Delta V$ is followed by a "=" or a "$\sim$" sign for spectroscopic and grism redshifts, respectively. From the top to the bottom, left to right:  4 objects show bridge or tidal features and the next 3 pairs show morphological peculiarities in both companions.  }}
\label{Companions}
\end{figure*}

 We searched for companions around the 41 galaxies within a projected separation $r_{proj}<$150~kpc, and rest-frame relative velocity $\Delta V<$ 500~km\,s$^{-1}$ for $z_{\rmn{spec}}$ \citep{2013MNRAS.433L..59P}, or $\Delta V<$1500~km\,s$^{-1}$ for $z_{\rmn{grism}}$. Amongst the 41 galaxies, 9 have neighbour(s) of comparable mass (J-band luminosity ratio $>$ 1:5) This number does not change much, increasing only to 11 or to 14, when accounting for $\Delta V$ limits of 3000 and 4500~km\,s$^{-1}$, respectively. This suggests that most of the pairs are physically linked rather than projections of galaxies occupying the same filament. Interestingly, 6 amongst the 9 pairs have another companion within $r_{proj}<$150~kpc and $\Delta V<$ 4500~km\,s$^{-1}$. Because $z_{\rmn{grism}}$ can have catastrophic misestimations, we also investigated whether projected close pairs that have failed the pair criterion in velocity could be physically linked. We found one object (3D-GS3-27242) in this case, which lies at only 13.5 kpc from another galaxy while the velocity difference is $\sim$ 6300~km\,s$^{-1}$ ($\Delta z_{\rmn{grism}}$). A closer look at HST images reveals a faint stellar bridge linking the two galaxies (S/N$>$3 in I-band, see Figure~\ref{Companions}).  \\

\begin{table*}
\centering
\caption{ Major merger candidates having a companion with a mass ratio above 0.15 and lying within 150~kpc and $\Delta V <$ 500 or 1500 ~km\,s$^{-1}$, depending on the origin of the redshift estimate (see their images in Figure~\ref{Companions}). (1) object name, (2) ID of the candidate in the 3D-HST catalogue \citep{2015arXiv151002106M}, (3) redshift of the candidate, (4) type of redshift measurement available for the companion (grism or spectroscopic observation), (5) stellar mass, (6) mass ratio between the object and the candidate, (7,8) projected distance $r_{proj}$ and velocity delta $\Delta v$ between the object and the candidate, (10) kinematic criterion not fulfilled, (11), morphological classification, (12) morpho-kinematic classification. Redshifts are extracted from the 3D-HST catalogue \citep{2015arXiv151002106M} and mass ratio from deep photometry in $(J+H)_{AB}$ bands. }
\begin{tabular}{l  c c cc  cc l c c c}
\toprule
   3D object & ID& z & z type  & Mass ratio & $r_{proj}$ & $ \Delta_V$& Comments & kinematic & morphology& \textbf{morpho-kin}.\\ 
     &  &     & &  & {\small(kpc)} & {\small(km/s)}& & criterion & class & class\\ 

   \midrule
COS3-6738  & 6550&0.870 & Spec &     0.33 &  64.22  &   256.96& Tidal features in both galaxies & 3,5&Pec & NV\\
COS3-13009&12401&0.954& Grism&0.27&61.67 &   456.70 &Tidal features in both galaxies & 3,4&Sp& SV\\
COS3-21583& 20612&0.903&Grism &0.55&146.66&151.42 &One disturbed galaxy & 3,4&Pec& NV\\
GS3-21045&31425&0.952 & Spec  &   0.86 &  37.73 &   460.35 & No morphological signatures &1,4 &Sp& SV\\
GS3-22005&33287 &0.955 & Spec  &   0.57 &  30.86 &  153.53 & Tidal features \& Bridge & 4&Pec& NV\\
GS3-27242 & 39825 & 1.068&Grism & 0.44      & 13.5  &  6300    & Bridge $^\dagger$ & --&Pec& SV\\
& &1.017 &Grism  &   0.44 &  67.70 &   1217.77  & & &\\
GS3-28388 &  41531&1.017&  Grism  &  0.63 &  38.50&    727.00 & One disturbed galaxy &-- & Pec& SV\\
		        &41775&1.019&  Grism  &  0.81 & 102.52&     378.34&  & &\\
U3-14150 & 19745&0.892 & Grism &   0.26 &  40.66  &  669.31& Two disturbed galaxies  & 3 & Pec& NV\\	        
U3-16817& 23207&0.788&  Grism   &  0.42 & 34.86&     409.85& Two disturbed galaxies & 5&Sp& SV\\ 
U3-25160&34315&0.898&  Grism  &   0.33 &  34.35 &   310.12& Two disturbed galaxies & --&Pec& SV\\
\bottomrule
\multicolumn{10}{l}{$^\dagger$ \textit{Possible redshift misidentification for this objet }}\\

\end{tabular}
\label{Table_interaction}
\end{table*}

The physical link between galaxies in the 10 detected close pairs is reasserted by the presence of tidal features, including bridges or tails, and/or peculiar morphologies.     
Table~\ref{Table_interaction} shows the properties of these 10 pairs: 4 pairs show evidence for tidal features or bridge, indicated by an arrow in Figure~\ref{Companions};  3 pairs are made of 2 galaxies having peculiar morphologies; and 2 pairs with a single galaxy being peculiar. This high number of tidal tails is surprising given the expected faintness of such features at high redshift, and may indicate the frequent occurrence of recent and strong interactions at z$\sim$1, at least in this sample. 

We found a good agreement between galaxies in close pairs and kinematic perturbations: 4 of the 6 galaxies that do not verify the rotating disk criterion (3) and 2 of the 5 galaxies that are affected by criterion (5) are in pairs.  In the same way, all 4 galaxies with PA misalignment are in pairs (see Figures~\ref{PA_mismatch} and ~\ref{Companions}). The two closest pairs, 3D-GS3-21045 and 22005\footnote{3D-GS3-21045 (3D-GS3-22005) are likely before (after) first passage from their regular spiral (perturbed) morphology, respectively.}, have their dynamical axes pointing towards their very close companions, as shown in Figure~\ref{PA_mismatch}. Such a behaviour is consistent with a merger in a phase before or just after a first passage. Finally, we do not observe kinematic perturbations in three pairs: 3D-GS3-27242, 28388 and 3D-U3-25160. This is consistent with the findings of \citet{2015ApJ...803...62H}:  by using redshifted templates of well defined local mergers, they showed that a significant fraction of the interacting disks and merger remnants are indistinguishable from isolated disks based on their sole kinematic classification. 

We found 10 (24\%) pairs amongst the 41 high S/N disk galaxies of W15, which include 6 non-virialized (75\%), 4 semi-virialized (29\%) and 0 virialized galaxies. This supports our analysis, and suggests that the high disk-fraction found by W15 is due to mis-identifications of galaxies that are instead experiencing a merger. However, the morpho-kinematic classification is meant to be conservative in how it secures the identification of isolated and virialized disks. It does not exclude the possibility that some of the semi-virialized galaxies could be rotating disk galaxies. The 25\% of isolated virialized spirals therefore corresponds to a lower limit of the fraction of disks at z$\sim$1, while at this stage (but see the discussion section), a strict upper limit is given by 53\% of galaxies dynamically classified as rotating disks.

\subsection{Merger rate}

In this section, we verify whether or not the estimated merger/disk fractions are consistent with a hierarchical mass assembly scenario. In \citet{2012ApJ...753..128P} an accurate calculation of the merger rate  was computed based on the number of objects observed in the different merger stages at $z\sim$0.6. This was done by systematically comparing the kinematics and morphologies of a representative sample of z$\sim$0.6 galaxies with hydrodynamical models of isolated disks and major mergers \citep{2009A&A...507.1313H}. The study presented hereafter only uses the galaxies in close pairs (see section \ref{pairs}). The lack of comparison with simulation to constrain the merger stages does not allow an accurate calculation of the merger rate for later merger phases (i.e., during and after fusion), as done in \citet{2012ApJ...753..128P}. 
 
The 10 pairs we identified are robust candidates for major mergers: 4 showing tidal features or bridges; 3 having both paired galaxies with peculiar morphologies expected after a 1st interaction; and one showing its kinematic axis pointing to its immediate neighbour (3D-GS3-21045).  The merger rate can only be calculated from the 72 galaxies defining the representative sample. In addition to the pairs detected amongst the 42 high S/N \textit{disk-like} galaxies, we had to include the number of pairs in the remaining 31 objects (15 low S/N disks and 16 non-disk galaxies). As mentioned in Section \ref{Sample}, we identified 24 objects over the 31, which were quasi equally distributed in non-disks (11) and low S/N disks (12) as in the parent sample. We assumed that they are sufficiently representative for a pair analysis.  Appendix~\ref{Pair_nd_lsd} shows the pair classification of the 23 galaxies using the same scheme as in Section~\ref{pairs}. We found 8 pairs, including 2 galaxies near fusion (3D-COS3-21481 and COS3-11933) and 4 galaxies showing tidal features (including a ring in 3D-COS3-08390), while all are showing severely distorted morphologies. Such a behaviour confirms the validity of the "non-disk" category of W15 that encompasses 6 out of the 8 pairs.

The fraction of pairs in the 72 galaxies sample is thus $f_\rmn{pairs}= (10 +8/23\times31)/72 =$29\% and the merger rate can be written as in \cite{2012ApJ...753..128P}:
\begin{equation}
r_\rmn{Merger}= 0.5 \, f_\rmn{pairs} \, f_\rmn{emi} \, / T_\rmn{prefusion} = 0.0685\,  \rmn{Gyr}^{-1}, 
 \label{rm}
\end{equation}
where 0.5 stands for the fact that only one galaxy in the pair is part of the sample and $f_{emi}$=0.85 is the fraction of emission line galaxies at $z\sim1$. It accounts for the fact that the observed sample included only emission-line galaxies \citep{1997ApJ...481...49H}. $T_{prefusion}=1.8\,$Gyr is the averaged timescale of the pre-fusion phase \citep[from observations and modelling, e.g.][ and references therein]{2012ApJ...753..128P}. \\

\begin{figure} 
\centering
 \resizebox{\hsize}{!}{\includegraphics{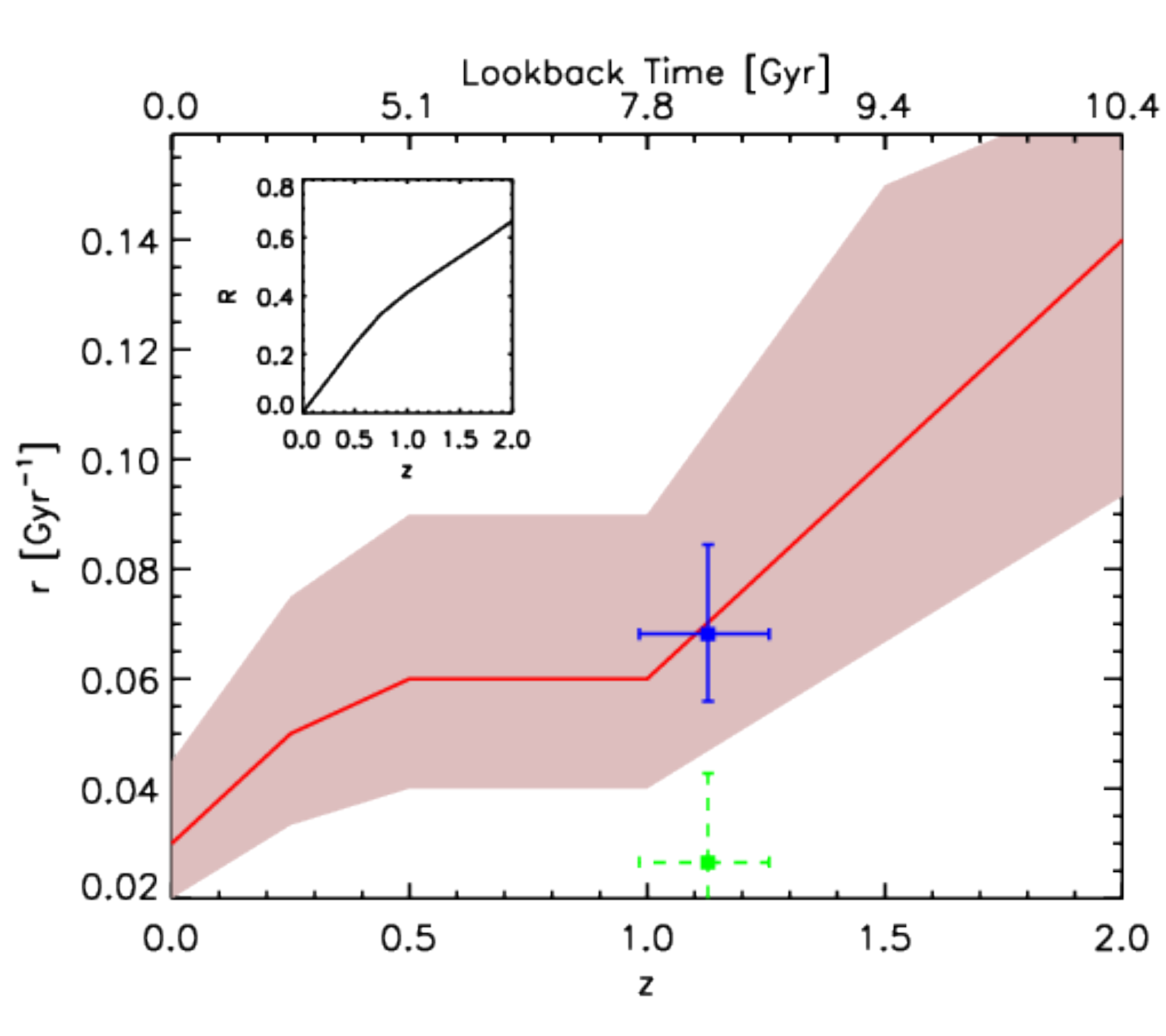}}
\caption{{\small  Galaxy merger rate, $r$, as a function of the redshift. The blue point gives the merger rate at the average pre-fusion phase epoch (z= 1.13), and has been calculated from the average redshift ($z_{mean}$=0.903) to which 1.8/2= 0.9~Gyr has been added. 
 The red curve shows the prediction of the semi-empirical
model \citep{2010ApJ...724..915H}, for the range of baryonic mass and mass fraction observed and the shaded red region shows the associated uncertainty.   
The red-shaded region shows the impact of possible systematic effects of the observational estimation of the merger
rate (see \citealt{2012ApJ...753..128P}). The green point gives the merger rate when only pair galaxies not identified as disks by W15 are taken into account.
The upper left inset shows the cumulative merger rate R as a function of redshift (i.e., from the full red curve of the main
panel).}}
\label{ratio_merger}
\end{figure}

Figure~\ref{ratio_merger} shows the resulting estimate of the merger rate for which
uncertainties were estimated as a quadratic combination of the
expected fraction of contaminations and the random uncertainties due to
statistical fluctuations in the sample, which were derived accurately
using confidence intervals for the binomial statistics. Theoretical
models usually give the merger rate at the time when halos start
merging, i.e., where galaxies that inhabit these halos are still in
pairs. The observed point in Fig.~\ref{ratio_merger}   is therefore shown at the median
redshift of the observed sample plus $0.5 \times T_{prefusion}$, which
is the average epoch at which the observed pairs were starting to
interact.

Figure~\ref{ratio_merger} also shows the prediction of the semi-empirical
$\Lambda$CDM model from \citet{2010ApJ...724..915H}, as used in \citet{2012ApJ...753..128P} to which we refer for details. In short, the predicted merger
rates were integrated over a varying range of baryonic mass $M_b$ as a
function of $z$ so that the observed range of $M_b$ at the median
redshift z$\sim$0.9 of the KMOS$^{3D}$ sample and their progenitors and
descendants at other epochs are probed. The resulting merger rates
were also integrated over the range of baryonic mass ratio 0.25-1.0 to
cover the range of mass ratios in the observed pairs (see Tab. 4).
This range was adopted as a definition of ``major mergers'' in the
following.

Figure~\ref{ratio_merger} indicates that the theoretical model is in a remarkable agreement with the
observed estimate without any fine-tuning, confirming the result of \citet{2012ApJ...753..128P} in a similar range of baryonic mass. The dominant source of
(random) uncertainty in the comparison is associated to the input
assumptions used in the model \citep{2010ApJ...724..915H}, which is illustrated
in Fig.~\ref{ratio_merger}  as a red region. These models, if not properly
scaled to the observed range of mass and mass ratio are typically
predictive within a factor $\sim$10 only \citep{hopkins09}. However,
if used in the exact same range of mass and mass ratios, they are accurate within a factor 2-3 compared to observations
 \citep{2012ApJ...753..128P}. For this reason, we refrained ourselves to compare
the present estimate with other determinations from the literature in
different ranges of mass and/or mass ratios, although we note that the
fit to the compilation of $M_{stellar}\sim10^{9.75}-10^{10.25}M_\odot$
galaxies\footnote{The authors quote a stellar mass range
$M_{stellar}\sim10^{10}-10^{10.50}M_\odot$ using a Salpeter IMF, which
we have converted into a Chabrier IMF as in Fig. 1 using a converting
factor $\sim$-0.25 dex.} as a function of redshift provided by \citet{2013A&A...553A..78L}
results in a major merger
rate $\sim$0.06 $Gyr^{-1}$ at z$\sim$ 1 (see their Eq. 9). Although their range of
stellar mass covers only partially the range of the present sample
(and especially the lowest mass bins, see Fig. 1), it is found in
quite good agreement with Fig.~\ref{ratio_merger}.

We also remark that such a fraction of major mergers is rather consistent with the succinct analysis of mergers in W15 (see their Section 4.2). They found 5 of them in very close pairs\footnote{besides 3D-COS3-21583 that we classified as non-virialized, we probably retrieved 3 of them in our analysis.}, separated by less than the IFU size. These very close pairs are observed during a very short period of the merging sequence, precisely near the first passage or just before the fusion. Using major merger hydro-dynamic simulations at different view angles \citep{2010ApJ...725..542H}, we estimated that the visibility window of these very close pairs is about 15\% of a major merger episode. When accounting for the short visibility window, the number identified by W15 is in fact consistent with the fraction of galaxies we found in other pre- or post-merger phases.

\section{Discussion and Conclusion}

A crucial test to decipher the main channel of spiral formation requires to estimate,  in the distant Universe, the fraction of isolated disks together with that of galaxies involved in a merger sequence. Fostered by the advent of new near-infrared IFS instruments, some studies focused on the presence of a rotation to estimate the disk fraction. However, rotation can be also induced by orbital motions in a merger sequence. In this study we investigated the limitations of kinematic and morpho-kinematic classifications to identify distant isolated disks, evidencing that the results depend on the number and on the detailed definition of the assumed selection criteria. Here we argue that rotating spirals are virialized systems having very specific characteristics, which can be detectable in high-z observations: 
\begin{enumerate}
\item The gas kinematics is dominated by rotation ($V/\sigma>$1, single velocity gradient). At low spatial resolution, the central velocity gradient translates into a peak in the velocity dispersion map;
\item In a virialized disk, gas and old stars are at equilibrium. The old stars and ionized gas disks rotate in the same plane and around the centre of mass of the system;
\item Disks have exponential light profiles and symmetric structures, e.g. spiral arms, bars, rings. Notice that the actual image depth in deep cosmological fields allows us to probe the $z\sim1$ disks up to 80\% of their optical radius. 
\end{enumerate}

Pure kinematic classifications based on only kinematic maps can only probe one aspect of what characterizes a spiral galaxy, and as such provide a crude upper limit of the disk fraction. The $\sigma$-centering criterion \citep{2006A&A...455..107F} is efficient in identifying perturbed rotations for distant galaxies, although it is mandatory to search perturbations over the whole dispersion map.  

Here we argue that it is equally essential to verify that gas and stars are at equilibrium for probing a virialized disk. Some of the high S/N disks of W15 show strong misalignments of their dynamical axes when compared to the optical ones, which is revealed after a proper measurement of the disk PAs (see, e.g., Figure~\ref{PA_mismatch}). In some others, the orbital motions of two closely interacting galaxies can be misinterpreted as a rotation \citep{2009A&A...496...51P, 2015ApJ...803...62H}. This is the case of 3D-COS-18471, which is rotationally-dominated but instead of being a rotating disk, it resembles a merger near fusion, explaining why the centre of rotation is well off the optical centre  (see Figure~\ref{Figure_high_SN_disk}). To identify these unrelaxed galaxies, two criteria imposing that gas and stars should share the same dynamical centre and PA have to be incorporated into the kinematic classification \citep{2008A&A...477..789Y}.

However, we show that even after incorporating 'gas/stars' equilibrium criteria, the kinematic classification still fails to identify all mergers. First, the coarse spatial resolution of kinematic observations can only detect extreme offsets between the kinematic and optical centers, generally at distances larger than the disk scale-length. Second, some of the criteria (3, 4 and 5) are not applicable to galaxies with extreme inclinations. For instance, measurements of PA are likely compromised in nearly face-on galaxies as well as detection of central $\sigma$-peaks because of the small velocity gradient projected in the centre. Moreover, disk galaxies in interactions maintain an ordered rotation during large time intervals, e.g., between first and second passages and during the relaxation phase (compare with NGC 5427/5426 and the M81 group). On the opposite, complex kinematics are visible during short periods of time, mainly during the first passage and coalescence phases \citep{2014MNRAS.443L..49P}.  A significant fraction of interacting disks and merger remnants are indistinguishable from isolated disks based on their sole kinematic classification \citep{2016ApJ...816...99H}. 

A multi-criteria approach is thus required to identify mergers in all phases from first passage to post-coalescence. This prompted us to define a morpho-kinematic classification \citep{2009A&A...507.1313H, 2016Book} that considers additional indicators such as the morphology to further identify virialized spirals. Imaging may reveal bridges and particularly tails in the outskirts of a galaxy, and since the inner parts come to equilibrium in a merger much faster than the outer parts, the kinematic analysis can misidentify it as a spiral in virialized equilibrium. The morpho-kinematic classification allows us to classify galaxies according to their degree of virialization (see Section \ref{Morpho-kine}), from virialized spirals to non-virialized systems. Applying the morpho-kinematic classification to the KMOS$^{3D}$ sample, we found that only 25\% of $z\sim1$ galaxies are virialized spirals, in contrast to the 80\% of \textit{disk-like} galaxies found by W15 in the same sample. 

Besides of being partly caused by the use of the morpho-kinematic classification in this paper, this discrepancy is also linked to the different methods used for evaluating morphological PAs and axis ratios, as well as to the use of numerous restrictions assumed by W15 (search for $\sigma$-peaks only along the dynamical axis, limiting the PA alignment to only b/a$<$ 0.6 galaxies) when deriving the kinematic classification\footnote{This paper methodology is further detailed in the handbook \citep{2016Book} that summarizes the numerous techniques - and their limitations - for selecting, observing, analysing and understanding distant galaxies.}.\\

To test the robustness of the morpho-kinematic classification for disentangling isolated disks from mergers, we confronted its results with an analysis of pairs. Using the open grism redshift survey 3D-HST, we found that 29\% of $z\sim1$ galaxies are physical pairs, including an almost similar fraction (24\%) within the high S/N disks of W15. All galaxies found in pairs are affected by either kinematic and/or morphological perturbations (classified as semi- or non-relaxed). Conversely, all galaxies classified as virialized spirals (VS) were found to be isolated. This consistency supports the validity of the morpho-kinematic classification. 

The pre-fusion time is approximatively equal to that of the post-fusion time \citep{2012ApJ...753..128P}.
Accounting for the visibility timescale of pairs, the 29\% of galaxies in pairs translate into 58\% of galaxies involved in a merger sequence at $z\sim1$. Morpho-kinematics is likely sensitive to perturbations for galaxies in pairs as well as to galaxies in fusion and post-fusion and leads to 75\% of semi or non-virialized galaxies. Combining the above arguments, it results that the fraction of isolated, virialized disks at z $\sim$ 1 ranges from 25 to 42\%, i.e., about one-third.  The merger fraction is also robustly established in close agreement with empirical, $\Lambda$CDM models. In contrast, the W15 kinematic classification leads to an inconsistent merger rate (see green point in Figure \ref{ratio_merger}). We conclude that most $z\sim1$ MW-mass field galaxies are involved into one or another phase of a major merger sequence, supporting the disk rebuilding scenario, and hence the hierarchical scheme of galaxy formation.

 \newpage
\setcounter{table}{0}
\begin{landscape}
 \begin{table}
\begin{center}
\caption{ Main properties:  Object name, redshift, stellar mass, half light radius from J-band, principal angle (PA) and b/a ratio measured from the J-H- summed image, Bulge-to-total ratio (B/T), morphological classification (classification flag between parenthesis: 0- all classifiers agree; 1- classifiers disagree), kinematic PA, separation in pixels between the optical and kinematic centre ($\Delta Centre$), difference between the morphological and kinematic PAs ($\Delta PA$), and $\Delta r_{kin}$. The next columns flag with $1$ the galaxies that verified kinematic criteria:  (1) Single velocity axis, (2) $V/\sigma >1$, (3) $\sigma$-peak coinciding  with rotation centre,  (4) $\Delta PA > 30^\degree$, and (5) Centre of rotation coinciding with optical centre. The last columns gives the Kinematic (classification flag between parenthesis, same as morphological classification) and the Morpho-Kinematic (M.K.) classifications. }
 \label{Table_param}
\begin{tabular}{lcccccccccc ccc rrrrr rrr}
\toprule
  \hline
  object & z & RA & DEC & LogMs & $R_{half}$ & $PA_{opt}$ & b/a & B/T & Morph. & $PA_{kin}$ & $\Delta c$ & $\Delta PA$& $\Delta r_{kin}$ &(1)&(2)&(3)&(4)&(5) & Kin & M.K. \\ 
            &     &       &        &{\small $(dex)$} &{\small (kpc)}      &{\small ($\degree$)}  &       &             &              & {\small ($\degree$)}&{\small (pixels)}&{\small ($\degree$)}&{\small ($\arcsec$)}&&&&&&& \\
\midrule
     \hline
 COS3\_10248 & 0.894 & 150.1379 & 2.2820 & 10.11 & 6.93 & 131 $\pm$ 7 & 0.51 $\pm$ 0.07 & 0.28 & Pec (0) & 139 & 0.07 & 8.00 & 0.39 & 0 & 1 & 0 & 1 & 1 & CK (0)& NV  \\ 
 COS3\_10857 & 0.952 & 150.1175 & 2.2863 & 9.93 & 3.65 & -91 $\pm$ 9 & 0.77 $\pm$ 0.09 & 0.52 & Pec (0) & -98 & 0.00 & 7.00 & 0.03 & 1 & 1 & 1 & 1 & 1 & RD (2)&SV  \\ 
 COS3\_13009 & 0.951 & 150.1004 & 2.3080 & 10.30 & 7.26 & -1 $\pm$ 12 & 0.84 $\pm$ 0.05 & 0.07 & Sp (0)  & 90 & 0.03 & 90.70 & 0.55 & 1 & 1 & 0 & 0 & 1 & CK (0)&  SV  \\ 
 COS3\_13311 & 0.893 & 150.1250 & 2.3103 & 9.74 & 5.53 & 0 $\pm$ 8 & 0.38 $\pm$ 0.10 & 0.00 & Sp (0)  & 0 & 0.02 & 0.30 & 0.07 & 1 & 1 & 1 & 1 & 1 & RD (0)&  RS  \\ 
 COS3\_14411 & 1.004 & 150.1367 & 2.3205 & 10.20 & 4.75 & -142 $\pm$ 7 & 0.42 $\pm$ 0.03 & 0.12 & Sp (1)  & -155 & 0.02 & 13.00 & 0.08 & 1 & 1 & 1 & 1 & 1 & RD (0)&  RS  \\ 
 COS3\_15061 & 0.900 & 150.1171 & 2.3271 & 10.19 & 3.72 & 149 $\pm$ 6 & 0.83 $\pm$ 0.05 & 0.00 & Sp (0)  & 141 & 0.00 & 8.08 & 0.02 & 1 & 1 & 1 & 1 & 1 & RD (0)&  RS  \\ 
 COS3\_16954 & 1.031 & 150.1033 & 2.3461 & 10.61 & 7.97 & -70 $\pm$ 5 & 0.58 $\pm$ 0.02 & 0.07 & Sp (0)  & -72 & 0.09 & 1.56 & 0.07 & 1 & 1 & 1 & 1 & 1 & RD (0)&  RS  \\ 
 COS3\_1705 & 0.827 & 150.0921 & 2.2044 & 10.35 & 7.60 & 73 $\pm$ 90 & 0.93 $\pm$ 0.08 & 0.12 & Sp  (0) & 106 & 0.00 & 32.84 & 0.04 & 1 & 1 & 1 & 1 & 1 & RD (0)&  RS  \\ 
 COS3\_18434 & 0.908 & 150.0717 & 2.3610 & 10.59 & 4.89 & 67 $\pm$ 90 & 0.91 $\pm$ 0.06 & 0.08 & Pec (1)  & 19 & 0.02 & 48.49 & 0.05 & 1 & 1 & 1 & 1 & 1 & RD (0)&  SV  \\ 
 COS3\_18471 & 0.900 & 150.1746 & 2.3608 & 10.24 & 4.54 & 121 $\pm$ 90 & 0.88 $\pm$ 0.10 & 0.00 & Pec (0)  & 3 & 3.00 & 63.00 & 0.21 & 1 & 1 & 1 & 1 & 0 & CK (0)&   NV  \\ 
 COS3\_19485 & 0.953 & 150.1592 & 2.3698 & 9.81 & 5.54 & 118 $\pm$ 3 & 0.53 $\pm$ 0.04 & 0.00 & Pec (0)  & 113 & 0.23 & 4.55 & 0.02 & 1 & 1 & 1 & 1 & 1 & RD (0)&  SV  \\ 
 COS3\_19935 & 0.953 & 150.1692 & 2.3746 & 10.08 & 5.15 & 13 $\pm$ 8 & 0.60 $\pm$ 0.03 & 0.07 & Pec (0) & 8 & 0.11 & 5.44 & 0.39 & 1 & 1 & 0 & 1 & 1 & PR (0)&   NV  \\ 
 COS3\_21583 & 0.902 & 150.1542 & 2.3912 & 10.46 & 6.54 & -64 $\pm$ 5 & 0.62 $\pm$ 0.02 & 0.24 & Pec (0) & -96 & 0.08 & 45.22 & 0.90 & 1 & 1 & 0 & 0 & 1 & CK (0)&   NV  \\ 
 COS3\_22796 & 0.914 & 150.0792 & 2.4054 & 10.27 & 9.29 & 44 $\pm$ 5 & 0.44 $\pm$ 0.01 & 0.04 & Pec (0)  & 57 & 0.02 & 12.67 & 0.06 & 1 & 1 & 1 & 1 & 1 & RD (0)&  SV  \\ 
 COS3\_23999 & 0.897 & 150.1463 & 2.4148 & 10.47 & 5.55 & 34 $\pm$ 90 & 0.88 $\pm$ 0.04 & 0.31 & Pec (0)  & 156 & 0.03 & 57.75 & 0.63 & 1 & 1 & 1 & 1 & 1 & RD (0)&   SV  \\ 
 COS3\_25038 & 0.852 & 150.1229 & 2.4267 & 10.78 & 8.06 & -68 $\pm$ 6 & 0.68 $\pm$ 0.05 & 0.19 & Sp (0)  & -57 & 0.05 & 11.18 & 0.08 & 1 & 1 & 1 & 1 & 1 & RD (0)&  RS  \\ 
 COS3\_26546 & 0.897 & 150.1533 & 2.4431 & 10.70 & 5.73 & 89 $\pm$ 90 & 0.92 $\pm$ 0.03 & 0.09 & Sp (1)  & 58 & 0.00 & 31.05 & 0.46 & 1 & 1 & 1 & 1 & 1 & RD (0)&  RV  \\ 
 COS3\_27071 & 1.029 & 150.0717 & 2.4475 & 9.91 & 3.73 & -43 $\pm$ 0 & 0.67 $\pm$ 0.02 & 0.00 & Pec  (0) & -51 & 0.08 & 7.89 & 0.02 & 1 & 1 & 1 & 1 & 1 & RD (0)&  SV  \\ 
 COS3\_4796 & 1.032 & 150.0733 & 2.2264 & 10.60 & 8.92 & 91 $\pm$ 0 & 0.62 $\pm$ 0.00 & 0.07 & Sp (1) & 90 & 0.00 & 0.92 & 0.02 & 1 & 1 & 1 & 1 & 1 & RD (0)&  RS  \\ 
 COS3\_5062 & 0.758 & 150.0904 & 2.2309 & 10.64 & 6.02 & -32 $\pm$ 11 & 0.84 $\pm$ 0.06 & 0.23 & Sp (0)  & -30 & 0.00 & 2.00 & 0.42 & 1 & 1 & 0 & 1 & 1 & PR (0)&  SV  \\ 
 COS3\_644 & 0.820 & 150.1287 & 2.1932 & 11.12 & 5.92 & -90 $\pm$ 2 & 0.65 $\pm$ 0.03 & 0.34 & Sp (0)  & -82 & 0.02 & 7.64 & 0.14 & 1 & 1 & 1 & 1 & 1 & RD (1)&  RS  \\ 
 COS3\_6511 & 0.803 & 150.1221 & 2.2435 & 10.43 & 3.13 & 108 $\pm$ 10 & 0.81 $\pm$ 0.04 & 0.58 & C (0)  & 126 & 0.00 & 18.13 & 0.05 & 1 & 1 & 1 & 1 & 1 & RD (0)&  SV  \\ 
 COS3\_6738 & 0.868 & 150.1677 & 2.2471 & 10.17 & 4.07 & -50 $\pm$ 5 & 0.77 $\pm$ 0.03 & 0.00 & Pec (0)  & -43 & 2.00 & 7.00 & 0.40 & 1 & 1 & 0 & 1 & 0 & CK (0)&   NV  \\ 
 GS3\_21045 & 0.955 & 53.1362 & -27.7632 & 11.09 & 7.01 & -37 $\pm$ 27 & 0.80 $\pm$ 0.04 & 0.27 & Sp  (0) & -110 & 0.00 & 73.00 & 0.20 & 0 & 1 & 1 & 0 & 1 & CK (0)&  SV  \\ 
 GS3\_22005 & 0.954 & 53.1246 & -27.7557 & 10.72 & 7.65 & -10 $\pm$ 6 & 0.72 $\pm$ 0.05 & 0.17 & Pec (0) & -64 & 0.03 & 54.43 & 0.03 & 1 & 1 & 1 & 0 & 1 & CK (0)&   NV  \\ 
 GS3\_23200 & 0.832 & 53.0458 & -27.7489 & 9.97 & 7.52 & -62 $\pm$ 90 & 0.92 $\pm$ 0.07 & 0.06 & Pec (0) & -7 & 0.05 & 55.36 & 0.08 & 1 & 1 & 1 & 1 & 1 & RD (0)&  SV  \\ 
 GS3\_27242 & 1.025 & 53.1308 & -27.7234 & 10.56 & 3.32 & 50 $\pm$ 90 & 0.90 $\pm$ 0.21 & 0.00 & Pec  (0) & 33 & 0.00 & 17.00 & 0.23 & 1 & 1 & 1 & 1 & 1 & RD (0)&  SV  \\ 
 GS3\_28388 & 1.022 & 53.0600 & -27.7163 & 9.99 & 3.65 & 167 $\pm$ 9 & 0.60 $\pm$ 0.03 & 0.00 & Pec  (0) & 148 & 0.13 & 18.76 & 0.10 & 1 & 1 & 1 & 1 & 1 & RD (0)&  SV  \\ 
 GS3\_30840 & 1.018 & 53.0533 & -27.7005 & 10.25 & 7.03 & -36 $\pm$ 90 & 0.93 $\pm$ 0.04 & 0.08 & Sp (1) & -75 & 0.00 & 39.04 & 0.52 & 1 & 1 & 1 & 1 & 1 & RV (0)&  RS  \\ 
 U3\_13321 & 0.912 & 34.3258 & -5.2155 & 10.85 & 4.02 & -162 $\pm$ 26 & 0.84 $\pm$ 0.08 & 0.41 & Sp (0) & 176 & 4.00 & 14.24 & 0.11 & 1 & 1 & 1 & 1 & 0 & CK (0)&  SV  \\ 
 U3\_14150 & 0.896 & 34.2417 & -5.2118 & 10.25 & 3.63 & 108 $\pm$ 20 & 0.88 $\pm$ 0.07 & 0.00 & Pec (0)  & 119 & 0.03 & 11.01 & 0.36 & 1 & 1 & 0 & 1 & 1 & PR (1)&   NV  \\ 
 U3\_15226 & 0.922 & 34.3383 & -5.2067 & 11.06 & 5.80 & 56 $\pm$ 2 & 0.83 $\pm$ 0.00 & 0.18 & Sp (0)  & 75 & 0.02 & 18.59 & 0.24 & 1 & 1 & 1 & 1 & 1 & RD (0)&  RS  \\ 
 U3\_16817 & 0.786 & 34.2875 & -5.2002 & 10.63 & 5.45 & 94 $\pm$ 4 & 0.57 $\pm$ 0.07 & 0.05 & Sp (0)  & 81 & 4.00 & 12.57 & 0.07 & 1 & 1 & 1 & 1 & 0 & CK (0)&  SV  \\ 
 U3\_18162 & 0.921 & 34.3579 & -5.1945 & 9.83 & 4.48 & 6 $\pm$ 6 & 0.74 $\pm$ 0.03 & 0.00 & Pec (0)  & 20 & 0.12 & 14.17 & 0.32 & 1 & 1 & 1 & 1 & 1 & RD (0)&  SV  \\ 
 U3\_18677 & 0.822 & 34.2912 & -5.1915 & 9.76 & 2.83 & -109 $\pm$ 10 & 0.67 $\pm$ 0.00 & 0.00 & C (0)  & -97 & 0.00 & 12.00 & 0.15 & 1 & 1 & 1 & 1 & 1 & RD (0)&  SV  \\ 
 U3\_25160 & 0.896 & 34.2696 & -5.1629 & 10.08 & 4.48 & -63 $\pm$ 8 & 0.76 $\pm$ 0.07 & 0.24 & Pec (1)  & -58 & 0.02 & 5.22 & 0.02 & 1 & 1 & 1 & 1 & 1 & RD (0)&  SV  \\ 
 U3\_3856 & 0.803 & 34.3592 & -5.2576 & 10.59 & 5.08 & -47 $\pm$ 6 & 0.82 $\pm$ 0.03 & 0.06 & Sp  (0) & -52 & 2.50 & 4.92 & 0.15 & 1 & 1 & 1 & 1 & 0 & CK (0)&  SV  \\ 
 U3\_4286 & 0.895 & 34.3038 & -5.2564 & 10.12 & 4.37 & -101 $\pm$ 6 & 0.82 $\pm$ 0.03 & 0.00 & Pec  (0) & -112 & 0.03 & 10.51 & 0.00 & 1 & 1 & 1 & 1 & 1 & RD (0)&  SV  \\ 
 U3\_5138 & 0.809 & 34.2496 & -5.2521 & 10.19 & 5.33 & -15 $\pm$ 4 & 0.71 $\pm$ 0.02 & 0.02 & Sp (0)  & -3 & 0.03 & 11.63 & 0.08 & 1 & 1 & 1 & 1 & 1 & RD (0)&  RS  \\ 
U3\_8072 & 0.824 & 34.3608 & -5.2392 & 10.68 & 3.37 & 49 $\pm$ 90 & 0.90 $\pm$ 0.03 & 0.24 & Sp (0)  & 20 & 0.00 & 29.38 & 0.05 & 1 & 1 & 1 & 1 & 1 & RD (0)&  RS  \\ 
 U3\_8493 & 0.786 & 34.3475 & -5.2369 & 10.70 & 3.04 & -1 $\pm$ 16 & 0.81 $\pm$ 0.05 & 0.61 & C (0)  & 20 & 0.03 & 22.00 & 0.00 & 1 & 1 & 1 & 1 & 1 & RD (1)&  SV  \\  
 \bottomrule
\end{tabular}
\end{center}
 \end{table}
\end{landscape}

\section*{Acknowledgements} 

We warmly thank the referee for her/his many insightful comments and suggestions which significantly contributed to improving the quality of the publication.  This work is based on observations taken by the 3D-HST Treasury Program (HST-GO-12177 and HST-GO-12328) with the NASA/ESA Hubble Space Telescope, which is operated by the Association of Universities for Research in Astronomy, Inc., under NASA contract NAS5-26555.

\bibliographystyle{mn2e}

\begin{thebibliography}{44}
\bibitem[\protect\citeauthoryear{Athanassoula et al.} {2016}]{2016arXiv160203189A} Athanassoula E., Rodionov S.~A., Peschken N., Lambert J.~C., 2016, arXiv, arXiv:1602.03189 
\bibitem[\protect\citeauthoryear{Aumer et al.}{2013}]{2013MNRAS.434.3142A} Aumer M., White S.~D.~M., Naab T., Scannapieco C., 2013, MNRAS, 434, 3142 
\bibitem[\protect\citeauthoryear{Barnes}{1988}]{1988ApJ...331..699B} Barnes J.~E., 1988, \apj, 331, 699 
\bibitem[\protect\citeauthoryear{Barrera-Ballesteros et al.}{2014}]{2014A&A...568A..70B} Barrera-Ballesteros, J.~K., Falc{\'o}n-Barroso, J., Garc{\'{\i}}a-Lorenzo, B., et al.\ 2014, \aap, 568, A70 
\bibitem[\protect\citeauthoryear{Barrera-Ballesteros et al.}{2015}]{2015A&A...582A..21B} Barrera-Ballesteros, J.~K., Garc{\'{\i}}a-Lorenzo, B., Falc{\'o}n-Barroso, J., et al.\ 2015, \aap, 582, A21 
\bibitem[\protect\citeauthoryear{Bellocchi et al.}{2012}]{2012A&A...542A..54B} Bellocchi, E., Arribas, S., \& Colina, L.\ 2012, \aap, 542, A54 
\bibitem[\protect\citeauthoryear{Brammer et al. }{2012}]{2012ApJS..200...13B} Brammer G.~B., et al., 2012, ApJS, 200, 13 
\bibitem[\protect\citeauthoryear{Chabrier}{2003}]{2003PASP..115..763C} Chabrier G., 2003, PASP, 115, 763 
\bibitem[\protect\citeauthoryear{Davies et al.}{2013}]{2013A&A...558A..56D} Davies, R.~I., Agudo Berbel, A., Wiezorrek, E., et al.\ 2013, \aap, 558, A56 
\bibitem[\protect\citeauthoryear{Dekel \& Birnboim}{2006}]{2006MNRAS.368....2D} Dekel A., Birnboim Y., 2006, MNRAS, 368, 2 
\bibitem[\protect\citeauthoryear{Delgado-Serrano et al.}{2010}]{2010A&A...509A..78D} Delgado-Serrano, R., Hammer, F., Yang, Y.~B., et al.\ 2010, \aap, 509, A78 
\bibitem[\protect\citeauthoryear{Dutton}{2009}]{Dutton09} Dutton, A.~A.\ 2009, MNRAS, 396, 121 
\bibitem[\protect\citeauthoryear{Epinat et al.}{2008}]{2008MNRAS.390..466E} Epinat, B., Amram, P., \& Marcelin, M.\ 2008, \mnras, 390, 466 
\bibitem[\protect\citeauthoryear{Epinat et al.}{2010}]{2010MNRAS.401.2113E} Epinat, B., Amram, P., Balkowski, C., \& Marcelin, M.\ 2010, \mnras, 401, 2113 
\bibitem[\protect\citeauthoryear{Faucher-Gigu{\`e}re, Kere{\v s}, \& Ma }{2011}]{2011MNRAS.417.2982F} Faucher-Gigu{\`e}re C.-A., Kere{\v s} D., Ma C.-P., 2011, MNRAS, 417, 2982 
\bibitem[\protect\citeauthoryear{Flores et al.}{2006}]{2006A&A...455..107F} Flores, H., Hammer, F., Puech, M., Amram, P., \& Balkowski, C.\ 2006, \aap, 455, 107 
\bibitem[\protect\citeauthoryear{Font et al. }{2011}]{2011MNRAS.416.2802F} Font A.~S., McCarthy I.~G., Crain R.~A., Theuns T., Schaye J., Wiersma R.~P.~C., Dalla Vecchia C., 2011, MNRAS, 416, 2802 
\bibitem[\protect\citeauthoryear{F{\"o}rster Schreiber et al.}{2006}]{2006ApJ...645.1062F} F{\"o}rster Schreiber, N.~M., Genzel, R., Lehnert, M.~D., et al.\ 2006, \apj, 645, 1062 
\bibitem[\protect\citeauthoryear{Genzel et al.}{2006}]{2006Natur.442..786G} Genzel, R., Tacconi, L.~J., Eisenhauer, F., et al.\ 2006, \nat, 442, 786 
\bibitem[\protect\citeauthoryear{Genzel }{2009}]{2009Natur.457..388G} Genzel R., 2009, Natur, 457, 388 
\bibitem[\protect\citeauthoryear{Gnerucci et al.}{2011}]{2011A&A...528A..88G} Gnerucci, A., Marconi, A., Cresci, G., et al.\ 2011, \aap, 528, A88 
\bibitem[\protect\citeauthoryear{Guedes et al. }{2011}]{2011ApJ...742...76G}  Guedes J., Callegari S., Madau P., Mayer L., 2011, \apj, 742, 76 
\bibitem[\protect\citeauthoryear{Gunn }{1982}]{Gunn82} Gunn, J. E., 1982, in Br¬uck H. A., Coyne G. V., Longair M. S., eds, Astrophysical Cosmology. Pontifical Scientific Academy, Vatican City, p. 233
\bibitem[\protect\citeauthoryear{Hammer et al.}{1997}]{1997ApJ...481...49H} Hammer, F., Flores, H., Lilly, S.~J., et al.\ 1997, \apj, 481, 49 
\bibitem[\protect\citeauthoryear{Hammer et al.}{2005}]{2005A&A...430..115H} Hammer, F., Flores, H., Elbaz, D., et al.\ 2005, \aap, 430, 115 
\bibitem[\protect\citeauthoryear{Hammer et al.}{2007}]{2007ApJ...662..322H} Hammer, F., Puech, M., Chemin, L., Flores, H., \& Lehnert, M.~D.\ 2007, \apj, 662, 322
\bibitem[\protect\citeauthoryear{Hammer et al.}{2009}]{2009A&A...507.1313H} Hammer, F., Flores, H., Puech, M., et al.\ 2009, \aap, 507, 1313 
\bibitem[Hammer et al.(2010)]{2010ApJ...725..542H} Hammer, F., Yang, Y.~B., Wang, J.~L., et al.\ 2010, \apj, 725, 542 
\bibitem[\protect\citeauthoryear{Hammer et al.}{2015}]{2015ApJ...813..110H} Hammer, F., Yang, Y.~B., Flores, H., Puech, M., \& Fouquet, S.\ 2015, \apj, 813, 110 
\bibitem[\protect\citeauthoryear{Hammer et al.}{2016}]{2016Book} Hammer, F.,  Puech, M., Flores, H., Rodrigues, M. 2016, Studying Distant Galaxies: a Handbook of Methods and Analyses, World Scientific, in press (ISBN: 978-1-78634-054-2)
\bibitem[\protect\citeauthoryear{Hopkins et al.}{2008}]{Hopkins08} Hopkins, P. F., Cox, T.~J., Hernquist, L.\ 2008, ApJ, 689, 17 
\bibitem[\protect\citeauthoryear{Hopkins et al.}{2009}]{2009ApJ...691.1168H} Hopkins, P.~F., Cox, T.~J., Younger, J.~D., \& Hernquist, L.\ 2009, \apj, 691, 1168 
\bibitem[\protect\citeauthoryear{Hopkins et al.}{2009}]{hopkins09} Hopkins, P.F. et al. 2009, \apj, 715, 202
\bibitem[\protect\citeauthoryear{Hopkins et al.}{2010}]{2010ApJ...724..915H} Hopkins, P.~F., Croton, D., Bundy, K., et al.\ 2010, \apj, 724, 915 
\bibitem[\protect\citeauthoryear{Hung et al.}{2015}]{2015ApJ...803...62H} Hung, C.-L., Rich, J.~A., Yuan, T., et al.\ 2015, \apj, 803, 62 
\bibitem[\protect\citeauthoryear{Hung et al.}{2016}]{2016ApJ...816...99H} Hung C.-L., Hayward C.~C., Smith H.~A., Ashby M.~L.~N., Lanz L., Mart{\'{\i}}nez-Galarza J.~R., Sanders D.~B., Zezas A., 2016, ApJ, 816, 99 
\bibitem[\protect\citeauthoryear{Husemann et al.}{2013}]{2013A&A...549A..87H} Husemann, B., Jahnke, K., S{\'a}nchez, S.~F., et al.\ 2013, \aap, 549, A87 
\bibitem[\protect\citeauthoryear{Kassin et al.}{2014}]{2014ApJ...790...89K} Kassin S.~A., Brooks A., Governato F., Weiner B.~J., Gardner J.~P., 2014, ApJ, 790, 89 
\bibitem[\protect\citeauthoryear{Kere{\v s} et al. }{2012}]{2012MNRAS.425.2027K} Kere{\v s} D., Vogelsberger M., Sijacki D., Springel V., Hernquist L., 2012, MNRAS, 425, 2027 
\bibitem[\protect\citeauthoryear{Koekemoer et al.}{2011}]{2011ApJS..197...36K} Koekemoer, A.~M., Faber, S.~M., Ferguson, H.~C., et al.\ 2011, \apjs, 197, 36 
\bibitem[\protect\citeauthoryear{Krajnovi{\'c} et al.}{2006}]{Krajnovic06} Krajnovi{\'c}, D., Cappellari, M., de Zeeuw, P.~T., \& Copin, Y.\ 2006, MNRAS, 366, 787 
\bibitem[\protect\citeauthoryear{Krist et al.}{2011}]{2011SPIE.8127E..0JK} Krist, J.~E., Hook, R.~N., \& Stoehr, F.\ 2011, \procspie, 8127, 81270J 
\bibitem[\protect\citeauthoryear{Law et al.}{2007}]{2007ApJ...669..929L} Law, D.~R., Steidel, C.~C., Erb, D.~K., et al.\ 2007, \apj, 669, 929 
\bibitem[\protect\citeauthoryear{L{\'o}pez-Sanjuan et al.}{2013}]{2013A&A...553A..78L} L{\'o}pez-Sanjuan, C., Le F{\`e}vre, O., Tasca, L.~A.~M., et al.\ 2013, \aap, 553, A78 
\bibitem[\protect\citeauthoryear{Lotz et al. }{2010}]{2010MNRAS.404..575L} Lotz J.~M., Jonsson P., Cox T.~J., Primack J.~R., 2010, MNRAS, 404, 575 
\bibitem[\protect\citeauthoryear{Mestel}{1963}]{Mestel63} Mestel L., 1963, MNRAS, 126, 553
\bibitem[\protect\citeauthoryear{Momcheva et al.}{2015}]{2015arXiv151002106M} Momcheva, I.~G., Brammer, G.~B., van Dokkum, P.~G., et al.\ 2015, arXiv:1510.02106 
\bibitem[\protect\citeauthoryear{Nelson et al.}{2013}]{2013MNRAS.429.3353N} Nelson D., Vogelsberger M., Genel S., Sijacki D., Kere{\v s} D., Springel V., Hernquist L., 2013, MNRAS, 429, 3353 
\bibitem[\protect\citeauthoryear{Patton et al.}{2013}]{2013MNRAS.433L..59P} Patton, D.~R., Torrey, P., Ellison, S.~L., Mendel, J.~T., \& Scudder, J.~M.\ 2013, \mnras, 433, L59 
\bibitem[\protect\citeauthoryear{Peng et al.}{2002}]{2002AJ....124..266P} Peng, C.~Y., Ho, L.~C., Impey, C.~D., \& Rix, H.-W.\ 2002, \aj, 124, 266 
\bibitem[\protect\citeauthoryear{Peirani et al.}{2009}]{2009A&A...496...51P} Peirani S., Hammer F., Flores H., Yang Y., Athanassoula E., 2009, A\&A, 496, 51 
\bibitem[\protect\citeauthoryear{Puech et al. }{2014}]{2014MNRAS.443L..49P} Puech M., Hammer F., Rodrigues M., Fouquet S., Flores H., Disseau K., 2014, MNRAS, 443, L49 
\bibitem[\protect\citeauthoryear{Puech et al.}{2012}]{2012ApJ...753..128P} Puech, M., Hammer, F., Hopkins, P.~F., et al.\ 2012, \apj, 753, 128 
\bibitem[\protect\citeauthoryear{Puech et al.}{2007}]{2007A&A...466...83P} Puech, M., Hammer, F., Lehnert, M.~D., \& Flores, H.\ 2007, \aap, 466, 83 
\bibitem[\protect\citeauthoryear{Shapiro et al.}{2008}]{2008ApJ...682..231S} Shapiro, K.~L., Genzel, R., F{\"o}rster Schreiber, N.~M., et al.\ 2008, \apj, 682, 231 
\bibitem[\protect\citeauthoryear{Skelton et al.}{2014}]{2014ApJS..214...24S} Skelton, R.~E., Whitaker, K.~E., Momcheva, I.~G., et al.\ 2014, \apjs, 214, 24 
\bibitem[\protect\citeauthoryear{Stott et al. }{2016}]{2016MNRAS.457.1888S} Stott J.~P., et al., 2016, MNRAS, 457, 1888 
\bibitem[\protect\citeauthoryear{Springel }{2010}]{2010ARA&A..48..391S} Springel V., 2010, ARA\&A, 48, 391 
\bibitem[\protect\citeauthoryear{Tomczak et al.}{2014}]{2014ApJ...783...85T} Tomczak, A.~R., Quadri, R.~F., Tran, K.-V.~H., et al.\ 2014, \apj, 783, 85 
\bibitem[\protect\citeauthoryear{Toomre \& Toomre }{1972}]{1972ApJ...178..623T} Toomre A., Toomre J., 1972, ApJ, 178, 623 
\bibitem[\protect\citeauthoryear{van der Wel et al.}{2012}]{2012ApJS..203...24V} van der Wel, A., Bell, E.~F., H{\"a}ussler, B., et al.\ 2012, \apjs, 203, 24 
\bibitem[\protect\citeauthoryear{van der Marel et al.}{2014}]{vanderMarel14} van der Marel, R. P., \& Kallivayalil, N G.\  2014, \apj, 781, 121
\bibitem[\protect\citeauthoryear{Vanzella et al.}{2008}]{2008A&A...478...83V} Vanzella E., et al., 2008, A\&A, 478, 83 
\bibitem[\protect\citeauthoryear{Vogelsberger et al.}{2014}]{Vogelsberger14} Vogelsberger, M., Genel, S., Springel, V. et al., 2014, Nature, 509, 177 
\bibitem[\protect\citeauthoryear{White \& Rees }{1978}]{1978MNRAS.183..341W} White S.~D.~M., Rees M.~J., 1978, MNRAS, 183, 341 
\bibitem[\protect\citeauthoryear{Whitaker et al.}{2014}]{2014ApJ...795..104W} Whitaker, K.~E., Franx, M., Leja, J., et al.\ 2014, \apj, 795, 104
\bibitem[\protect\citeauthoryear{Wisnioski et al.}{2015}]{2015ApJ...799..209W} Wisnioski, E., F{\"o}rster Schreiber, N.~M., Wuyts, S., et al.\ 2015, \apj, 799, 209 
\bibitem[\protect\citeauthoryear{Yang et al.}{2008}]{2008A&A...477..789Y} Yang, Y., Flores, H., Hammer, F., et al.\ 2008, \aap, 477, 789
\bibitem[\protect\citeauthoryear{Zheng et al.}{2004}]{2004A&A...421..847Z} Zheng, X.~Z., Hammer, F., Flores, H., Ass{\'e}mat, F., \& Pelat, D.\ 2004, \aap, 421, 847  
\end{thebibliography}

  \newpage
  
 \appendix

\section{High S/N disk from W15}
\label{Description_target}
This Annex provides the description of each object together with the arguments that led to the kinematic and morpho-kinematic classifications. The first acronym gives the morpho-kinematic class:  rotating spiral (RS); semi-virialized (SV); non-virialized (NV), and in parenthesis, the morphological class - spiral (Sp), peculiar (Pec), compact (C) - and the kinematic class - rotating disk (RD), perturbed rotation (PR), complex kinematics (CK).
\begin{description}
\item[\textbf{3D-COS3-644}:] RS (Sp/RD) - Spiral galaxy with slightly asymmetric residuals.  Rotating disk with a large contribution of the bulge to the large sigma peak. 
\item[\textbf{3D-COS3-1705}:] RS (Sp/RD) - Face-on spiral galaxy, with a bar detected in the residual map. The kinematics indicates a rotation.
\item[\textbf{3D-COS3-4796}:] RS (Sp/RD) - Spiral galaxy with a red bulge and a bar. The residual map reveals the spiral arms structure. The large clump at the left side could be a minor merger. The kinematics indicates a rotation.
\item[\textbf{3D-COS3-13311}:] RS (Sp/RD) - Spiral ring galaxy with symmetric residuals. The kinematics indicates a rotation.
\item[\textbf{3D-COS3-14411}:] RS (Sp/RD) - Spiral galaxy but unsecured classification. Could also be a peculiar: red core dominated galaxy  with a strong plume. The kinematics indicates a rotation with a large contribution of the bulge to the large sigma peak.
\item[\textbf{3D-COS3-15061}:] RS (Sp/RD) - Spiral. The object resembles a ring galaxy but the pseudo-ring is incomplete and highly asymmetric. Residuals are rather symmetric.  
\item[\textbf{3D-COS3-16954}:] RS (Sp/RD) - Sab type galaxy with symmetric residuals and with kinematics compatible with a rotation.
\item[\textbf{3D-COS3-25038}:] RS (Sp/RD) - Spiral galaxy with kinematics compatible with a rotation.
\item[\textbf{3D-COS3-26546}:] RS (Sp/RD) - Spiral galaxy, nearly face-on, with a possible tidal tail in the left side. The kinematic maps are consistent with a perturbed rotation with a $\sigma$-peak offset by 0.45\arcsec from the kinematic centre. iHowever, the lack of $\sigma$-peak at the centre of rotation could be due to the lack of $H\alpha$ emission in the bulge. This galaxy has therefore been classified as RD.
\item[\textbf{3D-GS3-30840}:] RS (Sp/RD) -  Spiral galaxy with a bar. The irregular morphology in J \& H  could be due to dust. The asymmetric residuals can be due to arms. There are three  $\sigma$-peaks, all of them offset from the kinematic centre. However, the lack of $\sigma$-peak at the centre of rotation could be due to the lack of $H\alpha$ emission in the bulge. This galaxy has therefore been classified as RD. 
\item[\textbf{3D-U3-5138}:] RS (Sp/RD) - Spiral galaxy with kinematics compatible with a rotating disk.
\item[\textbf{3D-U3-8072}:] RS (Sp/RD) - Face-on spiral galaxy with a ring visible in the residual maps and colour map. The kinematics indicates a rotation.
\item[\textbf{3D-U3-15226}:] RS (Sp/RD) - Spiral galaxy dominated by a bar (SBa). The kinematics indicates a rotation.
\end{description}

\begin{description}
\item[\textbf{3D-COS3-6511}:] SV (C/RD) - Compact galaxy ($R_{half}=3.13$kpc) with a ring detected in the residual map. The kinematics indicates a rotation.
\item[\textbf{3D-COS3-10857}:] SV (Pec/RD)  - Peculiar galaxy having two components with different colours. The residual map reveals asymmetric structures. The kinematics indicates a rotation.
\item[\textbf{3D-COS3-18434}:] SV (Pec/RD) - Almost face-on galaxy with peculiar morphology. There are strong asymmetries in the colour map. The structures in the residual map could be compatible with a ring but are rather asymmetric. The kinematics indicates a rotation.
\item[\textbf{3D-COS3-19485}:] SV (Pec/RD)- Peculiar galaxy with unsecured classification. The galaxy could be a spiral but is too asymmetric in H band. The residuals are also highly asymmetric. The kinematics is compatible with a rotation, although the $\sigma$-peak is very broad.
\item[\textbf{3D-COS3-22796}:] SV (Pec/RD) - Peculiar galaxy with a asymmetric low surface brightness disk. The elongated spot, a the left of the bulge, can be either a merger or a star-forming region. The residuals maps shows off-centered ring. The kinematics indicates a rotation.
\item[\textbf{3D-COS3-23999}:] SV (Pec/RD) - Peculiar galaxy with strong asymmetries. The nearby object is not at the same redshift. The kinematic maps suggest that the system is a perturbed rotation with a wide $\sigma$-peak located at 0.63\arcsec at the south of the bulge. However, the lack of $\sigma$-peak at the centre of rotation could be due to the lack of $H\alpha$ emission in the bulge. This galaxy has therefore been classified as RD. 
\item[\textbf{3D-COS3-27071}:] SV (Pec/RD) - Peculiar galaxy,  asymmetric or Tadpole. The kinematics indicates a rotation.
\item[\textbf{3D-GS3-23200}:] SV (Pec/RD) - Peculiar galaxy having strong irregularities at all wavelengths. Could be a merger close to completion (3rd passage, the 2nd galaxy almost destroyed in a polar encountering). The kinematics indicates a rotation.
\item[\textbf{3D-GS3-27242}:] SV (Pec/RD) - Peculiar galaxy. The galaxies is in interaction with a close companion. A bridge is visible between the two objects. The kinematics indicates a rotation.
\item[\textbf{3D-GS3-28388}:] SV (Pec/RD) - Peculiar galaxy with a V-shape in V band and a red centre. The centre is offset to the right when compared to the edge-on disk in J and H. The offset is confirmed by the residual maps. Nearby companion at the same redshift.The kinematics indicates a rotation.
\item[\textbf{3D-U3-18162}:] SV (Pec/RD) - Peculiar galaxy, with a tilted morphology with at least 2 components confirmed by the residual maps.The kinematics indicates a rotation.
\item[\textbf{3D-U3-18677}:] SV (C/RD) - Compact galaxy ($R_{half}=2.83$kpc) with kinematic compatible with a rotating disk. 
\item[\textbf{3D-U3-25160}:] SV (Pec/RD) - Peculiar galaxy with asymmetric residuals and small clumps. The bulge is offset from the centre of the disk component. The kinematics indicates a rotation.
\item[\textbf{3D-U3-4286}:] SV (Pec/RD) - Peculiar galaxy with a bar. The large clump is visible in all wavelength and could be a minor merger. The kinematics indicates a rotation.
\item[\textbf{3D-U3-8493}:] SV (C/RD) - Compact galaxy ($R_{half}=3.04$kpc) with a rotation.
\item[\textbf{3D-COS3-5062}:] SV (Sp/PR) - Spiral galaxy with a perturbed rotation. 
\item[\textbf{3D-COS3-13009}:] SV (Sp/CK) - Spiral galaxy with unsecured classification. The galaxy could be either a spiral with star forming clumps or minor merger, or a major merger in completion.  The mass ratio suggests a minor merger. The galaxy has a complex kinematics:  PAs are discrepant by $\sim90^\degree$. 3D-COS3-13009 is a candidate for a physical pair: there is a neighboring galaxy at 60\,kpc and 456\,km/s having tidal features. 
\item[\textbf{3D-GS3-21045}:] SV (Sp/CK) - Spiral galaxy dominated by a bar. The galaxy is separated by less than one optical radius from a companion at the same redshift. The galaxy has a complex kinematics: the velocity field has two gradients and the PAs are discrepant by $\sim70^\degree$. There are two $\sigma$-peaks with an offset of $0.2\arcsec$ from the centre of rotation, located at the extremity of the bar. 
\item[\textbf{3D-U3-3856}:] SV (Sp/CK) - Spiral galaxy. The centre of rotation is offset from the optical centre by 4 pixels. 
\item[\textbf{3D-U3-13321}:]  SV (Sp/CK) -  Spiral galaxy with a ring, a blue bar and a red bulge. The centre of rotation is offset from the optical centre by 2.5 pixels. 
\item[\textbf{3D-U3-16817}:] SV (Sp/CK) -  Edge-on galaxy with a complex centre which can be a bulge + ring spiral. The kinematics is complex because the centre of rotation is offset from the optical centre by 4 pixels. There is a companion at the same redshift separated by 34~kpc.
\end{description}

\begin{description}
\item[\textbf{3D-COS3-6738}:]  NV (Pec/CK) - Peculiar galaxy with asymmetric residuals. The galaxy has complex kinematics: the centre of rotation is located $\sim2$ pixels from the optical centre. The $\sigma$-peak is also offset by 0.40\arcsec from the rotation centre. 
\item[\textbf{3D-COS3-10248}:] NV (Pec/CK) - Peculiar galaxy with an extraordinary bright tidal tails (or gigantic arms?) and a bar. The object is probably relaxing after a very recent collision. The velocity field is complex, with two kinematic axes. The $\sigma$-peak is located at 0.39\arcsec to the north of the bulge.  
\item[\textbf{3D-COS3-18471}:] NV (Pec/CK) - Peculiar galaxy with an elongated red nucleus with a tidal tail. The object is probably an on-going or nearly complete merger. The kinematics is complex: the centre of rotation is located at 3 pixels east from the centre of rotation.  
\item[\textbf{3D-COS3-19935}:] NV (Pec/PR) - Peculiar galaxy, with strong  asymmetries in H Band and in the residuals map. Perhaps a minor merger with an edge-on Sp. The galaxy was classified as a perturbed rotation because the $\sigma$-peak is offset from the kinematic centre by 0.39\arcsec. 
\item[\textbf{3D-COS3-21583}:] NV (Pec/CK) - Peculiar galaxy with several components. The object is probably an on-going merger just before the fusion. The velocity maps are complex, with two kinematic axes. 
\item[\textbf{3D-GS3-22005}:] NV (Pec/CK) - Ring galaxy with a bridge linking it to the bottom-left galaxy. The object is part of a close pair (31kpc), probably an on-going merger after the 1st passage. The residual-map shows asymmetric spiral arms and the bridge connecting the two galaxies. The kinematics is complex: the PAs mismatch $\sim54^\degree$. The kinematic PA is oriented toward the companion. 
\item[\textbf{3D-U3-14150}:] NV (Pec/PR) - Peculiar galaxy with a blue core and complex residuals. The object is probably a merger near fusion. There is a companion 40$\,$kpc away. 
\end{description}

\begin{figure} 
\caption{From left to right:  i+J+H colour images, (i--J) colour map, {\sc Galfit} residuals and decomposition of the major axis light profile for the 41 galaxies classified in this work. Galaxies are ordered as in Appendix A, according to their morpho-kinematic classification: virialized spirals, semi-virialized (rotating disk+peculiar morphology, perturbed kinematics + spiral morphology) and non-virialized. The morphological and kinematic classification are indicated in the color image. The later also indicates the  optical centre as a pink cross, the morphological PA as a pink line, the centre of rotation and  kinematic PA are respectively the cyan cross and cyan dashed line, and the peaks of dispersion are delimited by a cyan square.. Notice that when the optical and kinematic centers coincide, the optical centre (pink cross) is barely visible. The two panels on the left are the residuals and then the fit from the {\sc galfit} decomposition.  When the galaxy is found in a pair a wide field is shown, indicating the redshift of the companions in the field. \textbf{Figure available in:} \url{http://mygepi.obspm.fr/~mrodrigues/Doc/KMOS-3DHST_final.pdf}. }
\label{Figure_high_SN_disk}
\end{figure}

\section{The 23 identified non-disk (ND) and low S/N disks (LSD) of W15}

\begin{figure*}
\centering
\caption{{\small  Eight galaxies in pairs within  $r_{proj}<$150~kpc and $\Delta V< 1500$~km$s^{-1}$ when calculated from $z_{\rmn{grism}}$ (otherwise $\Delta V< 500$~km$s^{-1}$). In the different panels, $\Delta V$ is followed by a "=" or a "$\sim$" sign for spectroscopic and grism redshifts, respectively. \textbf{Figure available in:} \url{http://mygepi.obspm.fr/~mrodrigues/Doc/KMOS-3DHST_final.pdf}. 
 }}
\label{Pair_nd_lsd}
\end{figure*}

\end{document}